\documentclass[twocolumn,preprintnumbers,amsmath,amssymb,superscriptaddress,nofootinbib,longbibliography]{revtex4-1}

\usepackage{graphicx}
\usepackage{bm}
\usepackage{dsfont}
\usepackage[usenames,dvipsnames]{xcolor}
\usepackage{pstricks}
\usepackage[tight]{subfigure}
\usepackage{verbatim}
\usepackage{units}
\usepackage{multirow}
\usepackage{enumitem}
\usepackage{mathrsfs}
\usepackage{leftidx}
\usepackage{xspace}
\usepackage{pifont}

\usepackage[a4paper]{hyperref}
\hypersetup{colorlinks=true,linktoc=all,linkcolor=blue,breaklinks=true,citecolor=blue,urlcolor=blue}

\usepackage[english]{babel}
\newcommand{\intd}{\text{d}}

\newcommand{\kB}{k_\text{B}}

\newcommand{\bra}[1]{\langle #1|}
\newcommand{\ket}[1]{|#1\rangle}
\newcommand{\brav}[1]{{\langle}#1|}
\newcommand{\ketv}[1]{{|#1\rangle}}

\newcommand{\veco}[1]{ \hat{\bm{#1}}}

\newcommand{\Ham}{\skew{3}{\hat}{\mathcal{H}}}
\newcommand{\opS}{\skew{3}{\hat}{S}}

\newcommand{\opn}{\skew{1}{\hat}{n}}
\newcommand{\opa}{\skew{1}{\hat}{a}}
\newcommand{\opb}{\skew{1}{\hat}{b}}

\newcommand{\opAcv}{\hat{\mathcal{A}}_\text{c}}
\newcommand{\opAsv}{\hat{\mathcal{A}}_\text{s}}
\newcommand{\opSV}{\skew{2.5}{\hat}{\mathcal{S}}}
\newcommand{\opSVnq}{\skew{2.5}{\hat}{\mathcal{S}}_{\!nq}}

\renewcommand{\lq}{\lambda_q}
\newcommand{\Luqn}{\Lambda^\text{u}_{nq}}
\newcommand{\Ltqn}{\Lambda^\text{t}_{nq}}
\newcommand{\Lu}{\Lambda^\text{u}}
\newcommand{\Lt}{\Lambda^\text{t}}
\newcommand{\Lucrit}{\Lambda^\text{u}_{1,\text{crit}}}

\newcommand{\Hmol}{\skew{3}{\hat}{\mathcal{H}}_\text{mol}}

\newcommand{\Hjun}{\skew{3}{\hat}{\mathcal{H}}_\text{jun}}
\newcommand{\Hg}{\skew{3}{\hat}{\mathcal{H}}_\text{ch}}
\newcommand{\Htun}{\skew{3}{\hat}{\mathcal{H}}_\text{tun}}
\newcommand{\Hspin}{\skew{3}{\hat}{\mathcal{H}}_{\text{spin}}}
\newcommand{\Hvib}{\skew{3}{\hat}{\mathcal{H}}_{\text{vib}}}
\newcommand{\Hchvib}{\skew{3}{\hat}{\mathcal{H}}_{\text{ch-vib}}}
\newcommand{\Hspinvib}{\skew{3}{\hat}{\mathcal{H}}_{\text{spin-vib}}}

\newcommand{\opd}{\skew{2}{\hat}{d}}

\newcommand{\Vg}{V_\text{g}}
\newcommand{\Vb}{V_\text{b}}

\newcommand{\nv}{n_\text{v}}

\newcommand{\en}{\mathcal{E}}
\newcommand{\tunT}{\mathcal{T}^\sigma}
\newcommand{\FCcoef}{\mathcal{J}}

\newcommand{\Dnt}{\delta D_n^{(2)}}
\newcommand{\Dnf}{\delta D_n^{(4)}}
\newcommand{\Ent}{\delta E_n^{(2)}}
\newcommand{\Enf}{\delta E_n^{(4)}}
\newcommand{\Cnf}{\delta C_n^{(4)}}
\newcommand{\Deff}{D_\text{eff}}
\newcommand{\Eeff}{E_\text{eff}}

\newcommand{\coefC}{\mathcal{C}}

\newcommand{\rhored}{\skew{0.5}{\hat}{\varrho}^\text{red}}
\newcommand{\rhoP}{\mathcal{P}}
\newcommand{\rhoPog}{\mathcal{P}_{1\text{g}}}

\newcommand{\og}{{1\text{g}}}
\renewcommand{\oe}{{1\text{e}}}
\newcommand{\coh}{\text{coh}}

\newcommand{\Chi}{\psi}

\newcommand{\vSigmaog}{\bm{\Sigma}_{1\text{g}}}
\newcommand{\Sigmaog}{\Sigma_{1\text{g}}}

\newcommand{\magB}{\bm{\mathcal{B}}}
\newcommand{\trW}{\mathcal{W}}

\newcommand{\etal}{\emph{et al.}\xspace}
\newcommand{\via}{\emph{via}\xspace}
\newcommand{\ie}{\emph{i.e.}\xspace}

\newcommand{\eg}{\emph{e.g.}\xspace}

\newcommand{\circon}{\protect\raisebox{-1.25pt}{\protect\scalebox{1.3}{\ding{192}}}\xspace}
\newcommand{\circtw}{\protect\raisebox{-1.25pt}{\protect\scalebox{1.3}{\ding{193}}}\xspace}
\newcommand{\circth}{\protect\raisebox{-1.25pt}{\protect\scalebox{1.3}{\ding{194}}}\xspace}
\newcommand{\circfo}{\protect\raisebox{-1.25pt}{\protect\scalebox{1.3}{\ding{195}}}\xspace}
\newcommand{\circfi}{\protect\raisebox{-1.25pt}{\protect\scalebox{1.3}{\ding{196}}}\xspace}
\newcommand{\circsi}{\protect\raisebox{-1.25pt}{\protect\scalebox{1.3}{\ding{197}}}\xspace}

\usepackage[normalem]{ulem}

\begin{document}


\title{Vibration-induced modulation of magnetic anisotropy in a magnetic molecule}

\author{Ahmed Kenawy}
\affiliation{Institute for Theoretical Physics, KU Leuven, B-3001 Leuven, Belgium}
\affiliation{Department of Microtechnology and Nanoscience MC2, Chalmers University of Technology, SE-412 96 G\"{o}teborg, Sweden}
\author{Janine Splettstoesser}
\affiliation{Department of Microtechnology and Nanoscience MC2, Chalmers University of Technology, SE-412 96 G\"{o}teborg, Sweden}
\author{Maciej Misiorny}
\email{misiorny@amu.edu.pl}
\affiliation{Department of Microtechnology and Nanoscience MC2, Chalmers University of Technology, SE-412 96 G\"{o}teborg, Sweden}
\affiliation{Faculty of Physics, Adam Mickiewicz University, PL-61 614 Pozna\'{n}, Poland}

\date{\today}

\begin{abstract}
We  theoretically analyze the spectrum of a magnetic molecule when its charge and spin can couple to the molecular vibrations.
More specifically, we show that the interplay between charge-vibron and spin-vibron coupling leads to a renormalization of the magnetic anisotropy parameters of the molecule. 
This effect is discussed for a model  device consisting of an individual magnetic molecule embedded in a  junction. We study the transport properties of the device and illustrate how the differential conductance is affected by the vibrationally induced renormalization of the magnetic anisotropy.
Depending on the total molecular spin and the bare (intrinsic) magnetic anisotropy, the induced modulation can lead to visible shifts and crossings in the spectrum, and it can even be the cause of a transport blockade.
It is therefore of particular interest to use mechanically controllable break junctions, since in such a case, the relevant coupling between the molecular spin and vibrations can be \textit{controlled} \via deformations of the molecule when stretching or compressing the junction.
\end{abstract}

\maketitle

\section{Introduction}
\label{sec:introduction}

Interest in molecular electronics~\cite{Xiang2016} is stimulated by rapid technological advances that allow for isolation and manipulation of individual molecules to realize single-molecule junctions~\cite{Perrin2015Feb,Huang2015Feb} ---nanoscopic devices with tunable optical, mechanical and magnetic properties~\cite{Tao2006}. 
One particularly prospective candidate for information storing and processing devices are molecules that exhibit large effective spin and magnetic anisotropy. The combination of these two quantities gives rise to magnetic bistability, which is a key prerequisite for a system to serve as a memory element~\cite{Bartolome_book}. Accordingly, the control of the magnetic anisotropy of molecules deposited in a junction is imperative for achieving functional devices. 
So far, only a few schemes for modifying such magnetic anisotropy \emph{in situ} have been demonstrated experimentally in specific molecules. For instance, by means of electrical gating, dissimilar magnetic properties of different molecular charge states were utilized~\cite{Zyazin2010}, or, by mechanical straining of the junction, the ligand field in a molecule based on a single magnetic ion was locally altered~\cite{Parks2010}. 
In addition, theoretical analysis predicts that  also  application of effective spintronic fields should be a feasible approach~\cite{Misiorny2013Dec}.
In this paper, we explore another possible way of engineering magnetic anisotropy in large-spin molecules which harnesses the coupling between spin and molecular vibrations without the application of external fields to the molecule.  

Individual molecules inserted in junctions vibrate with discrete frequencies, and these quantized vibrations (so-called \emph{vibrons}) can couple to other molecular  degrees of freedom, such as, charge and spin. 
For example, the interaction between electronic charge and vibrations can lead to excitation of transitions between different molecular vibrational states, when an electron tunnels through a molecule. This effect has been experimentally observed in single-molecule junctions based on carbon derivatives, specifically carbon nanotubes and fullerenes~\cite{LeRoy2004,Pradhan2005,Pasupathy2005,Sapmaz2006,Leturcq2009,Benyamini2014}, and also in other single molecules~\cite{Stipe1998,Yu2002,Leon2008,Osorio2010,Franke2012}. 
Moreover, if this \emph{charge-vibron} coupling is strong, it drastically impacts the transport properties of individual molecules, and at low bias-voltage it may even block transport of electrons ---an effect known as Franck Condon blockade~\cite{Koch2005,Koch2006Nov}. Recently, this effect has been experimentally and theoretically studied also in the context of magnetic molecules~\cite{Burzur2014,McCaskey2015}.
On the other hand, the primary interest in the coupling between \emph{vibrations} and \emph{spins} stems from its prominent role in the spin relaxation processes, which have been extensively studied for various  systems, \eg, atomic spins in crystal solids~\cite{Orbach1961Dec,Stoneham1965} and other molecular systems~\cite{Villain1994,Leuenberger1999,Garanin1997,Chudnovsky2005,Park2008,Kokado2010Spin}.
However, only recently, the effect of \emph{spin-vibron} coupling on the properties of individual molecules captured in junctions has caught some attention~\cite{May2011,Ruiz2012}. 
It has been suggested for sensing~\cite{Ohm2012,Palyi2012May} and cooling~\cite{Stadler2014,Stadler2015} applications in carbon nanotubes, and experimentally demonstrated to arise between a single molecular spin and a carbon nanotube~\cite{Ganzhorn2013}.

Here, we address the general question of how the interplay of the charge- and spin-vibron coupling in a single magnetic molecule affects its magnetic properties. While in this paper we deal with a general model that could be relevant for a large class of molecules, we would like to point out that the influence of (static) deformations on the magnetic anisotropy has recently been experimentally demonstrated in Co-based molecules~\cite{Parks2010}. 
For the purpose of this paper, we consider a model device consisting of a spin-anisotropic molecule embedded in a molecular junction, where vibrations of the molecule couple to both, the charge of tunneling electrons and the resulting spin of the molecule. 
To analyze the effect of vibrations on magnetic properties of the molecule, we derive an \textit{effective} giant-spin Hamiltonian exhibiting relevant corrections to the magnetic anisotropy constants due to the charge- and spin-vibron coupling. We show that such corrections significantly impact the spectral properties of the molecule, which, in turn, can have a profound effect on transport characteristics of the device. 
In particular, we here analyze signatures in the differential conductance emerging from the modulation of the magnetic anisotropy of the molecule due to the interplay of charge- and spin-vibron couplings. 
In order to calculate transport properties of the weakly coupled molecule, we use a master equation approach deriving from a real-time diagrammatic technique. An additional \textit{technical} achievement of this paper is the careful analysis of the regimes where coherent superpositions of molecular states do not affect the transport properties. We  thereby validate the simpler master equation approach, where such superpositions are disregarded,  for the situations studied here. 

This paper is organized as follows: the model of a vibrating magnetic molecule captured in a three-terminal molecular junction  is introduced in Sec.~\ref{sec:theory}, whereas the effective spin Hamiltonian including corrections to magnetic anisotropy constants due to the charge- and spin-vibron couplings is derived in Sec.~\ref{sec:states_and_effHams}. 
Next, in Sec.~\ref{sec:spectrum} we discuss how these couplings affect spectral properties of the molecule.
Key transport characteristics of this system
are presented in Sec.~\ref{sec:transport}. 
%
Finally, a summary of the main findings and conclusions are given in Sec.~\ref{sec:conclusions}.  Appendix~\ref{app:coherences} contains an analysis of the role of coherent superpositions between molecular states for transport calculations.

\section{Model of a vibrating magnetic molecule in a magnetic junction}
\label{sec:theory}

\begin{figure}[t!]
	\includegraphics[width=0.85\columnwidth]{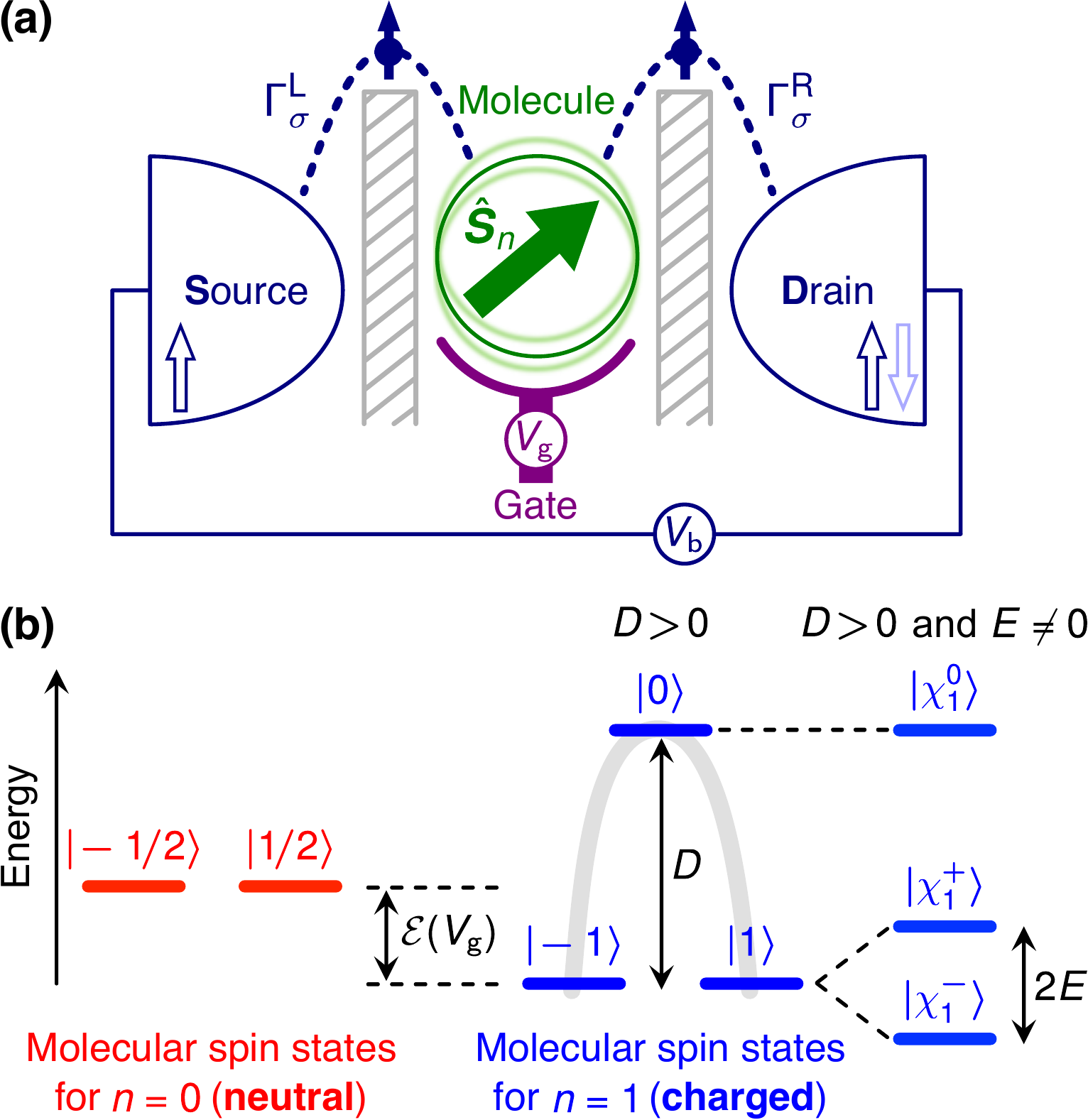}
	\caption{%
	(a) Schematic illustration of a single magnetic molecule (represented as an effective spin~$\veco{S}_n$) embedded between two  ---possibly magnetic--- electrodes, with collinear (parallel or \mbox{antiparallel}) configuration of their spin moments.
	A gate electrode is used to tune the energy spectrum of the charged molecule.
	(b)~Effect of magnetic anisotropy~on the spectrum of a model molecule with spins~\mbox{$S_0=1/2$} and \mbox{$S_1=1$}, given by   spin states \mbox{$\ket{\Chi_0}\in\big\{\ket{\pm 1/2}\big\}$} for the neutral state and~\mbox{$\ket{\Chi_1}\in\big\{\ket{0},\ket{\pm1}\big\}$} for the charged state with uniaxial anisotropy only (\mbox{$E=0$}). In the presence of transverse anisotropy (\mbox{$E\neq0$}), we get
	$
		\ket{\Chi_1}
		\in
		\big\{
		\mbox{$
		\ket{\chi^0_1}
		\equiv
		\ket{0}$},
		\mbox{$\ket{\chi_1^\pm}$}
		\equiv
		\big(\mbox{$\ket{1}$}\pm\mbox{$\ket{-1}$}\big)/\sqrt{2}
		\big\}
	$.  
	For further explanation see Sec.~\ref{sec:model_mol}.
	}
	\label{fig1}
\end{figure}

In this section, we formulate the model for a magnetic molecule embedded in a junction, as depicted in Fig.~\ref{fig1}(a). The key features of such a model are captured by the general  Hamiltonian
\begin{equation}\label{eq:H_tot}
	\Ham
	=
	\Hmol
	+
	\Hvib
	+
	\Hjun
	.
\end{equation}
Importantly, the characteristics of a molecule are typically strongly impacted by its vibrational degrees of freedom. Only, when introducing the model, for conceptual clarity, we formally split the  part of the Hamiltonian corresponding to the molecule, \mbox{$\Hmol+\Hvib$}, into two parts:
(i) $\Hmol$ describing the charge and spin properties of a static molecule (see Sec.~\ref{sec:model_mol}), and 
(ii) $\Hvib$ including the effects associated with molecular vibrations (see Sec.~\ref{sec:model_vib}). 
Finally, the last term of Eq.~(\ref{eq:H_tot}), $\Hjun$, accounts for the bare magnetic junction as well as for tunneling of electrons between electrodes of the junction and the molecule (see Sec.~\ref{sec:model_tun}).

\subsection{Magnetic molecule}
\label{sec:model_mol}

We consider a class of magnetic molecules whose static properties are determined by their charge and spin states. The associated energy is described by the Hamiltonian
\begin{equation}\label{eq:H_mol}
	\Hmol
	=
	\Hg
	+
	\Hspin
	.
\end{equation}
The first term of the Hamiltonian above arises due to the capacitive coupling of the molecule to a gate voltage~$\Vg$, which shifts the entire spectrum of the  molecule by an energy~\mbox{$\propto e\Vg$} depending on its charge. Specifically, we assume that only two charge states~$n$ of the molecule are energetically accessible: the \emph{neutral} state~(\mbox{$n=N$}) and the \emph{charged} state~(\mbox{$n=N+1$}). For notational brevity we henceforth set~$N$ to 0.
In principle, the occupation of many different molecular orbitals can lead to these two charge states; the occupation number operator of the molecule therefore reads as
\mbox{$
	\opn
	\equiv
	\sum_{l,\sigma}
	\opd_{l\sigma}^\dagger\opd_{l\sigma}^{}
	,
$}
with~$\opd_{l\sigma}^\dagger\,(\opd_{l\sigma}^{})$ standing for the operator creating (annihilating) a spin-$\sigma$ electron in the $l$th molecular 
orbital.\footnote{%
Note that the operator~$\opn$ is formally defined as \mbox{$\opn-N$}, that is, it counts only the number of excess electrons with respect to the neutral charge state.}
Consequently, the effect of capacitive coupling of the molecule to a gate electrode is simply given by~\mbox{$\Hg=\en(\Vg)\opn$}, with a gate-voltage dependent energy~$\en$.

From the magnetic point of view, in each charge state~$n$ the molecule can be regarded as an effective ground-state \emph{molecular} spin~$\veco{S}_n$, whose intrinsic magnetic behavior is characterized by the giant-spin Hamiltonian~\cite{Kahn_book,Gatteschi_book},
\begin{equation}\label{eq:H_spin}
	\Hspin 
	= 
	\sum_{n=0,1}
	\!
	\Big\{
	\!
	-D_n
	\big(\opS_n^z\big)^{\!2}
	+
	E_n
	\Big[
	\big(\opS_n^x\big)^{\!2}
	-
	\big(\opS_n^y\big)^{\!2}
	\Big]
	\Big\}
	.
\end{equation}
In the equation above, the first term represents the \emph{uniaxial} component of the magnetic anisotropy, while the \emph{transverse} component is described by the second term. The relevant anisotropy constants are given by~$D_n$ and~$E_n$. 
This magnetic anisotropy can, for instance, stem from a static deformation of the molecule due to the deposition into the junction.

In order to gain insight about the magnetic behavior of the static model molecule, it is instructive to analyze the eigenstates of the Hamiltonian as given in Eq.~(\ref{eq:H_mol}), 
\mbox{$
	\ket{\Chi_n}
$,}
with 
\mbox{$
	\Hmol\ket{\Chi_n}
	=
	\en_{\Chi_n}\ket{\Chi_n}
$.}
In the situation when a molecule exhibits exclusively a uniaxial component of magnetic anisotropy (\mbox{$D_n\neq0$} and~\mbox{$E_n=0$}), the basis of eigenstates of the molecule is simply formed by the states~\mbox{$\big\{\ket{\Chi_n}\equiv\ket{S_n,M_n}\big\}$} representing projections of the spin~$\veco{S}_n$ on the $z$-axis, that is, 
\mbox{$
	\opS_n^z\ket{S_n,M_n}
	=
	M_n\ket{S_n,M_n}
$}.
Note that  a~magnetic molecule in a given charge state can in general exhibit a~few spin multiplets (with different total spin~$S_n$). These spin multiplets are typically very well separated in energy, so that only states belonging to the ground spin multiplet are energetically accessible in the parameter regime under consideration. Therefore, in the following we often use a simplified notation replacing \mbox{$\ket{S_n,M_n}\rightarrow\ket{M_n}$}.  
Now, if also the transverse component is present (\mbox{$D_n\neq0$} and~\mbox{$E_n\neq0$}), one finds that the eigenstates~$\big\{\ket{\Chi_n}\big\}$ correspond to linear combinations of the spin projections along the~$z$-axis, given by 
\mbox{$
	\ket{\Chi_n}
	=
	\sum_{M_n}
	\coefC_{M_n}^{\Chi_n}
	\ket{M_n}
$},
where $\coefC_{M_n}^{\Chi_n}$ are the expansion coefficients.

To~illustrate the effect of magnetic anisotropy on the energy spectrum of a molecule in a given vibrational state, in~Fig.~\ref{fig1}(b) we show the energy spectrum for a hypothetical molecule with~\mbox{$S_0=1/2$} and~\mbox{$S_1=1$}, additionally assuming that \mbox{$D_0=E_0=0$}, \mbox{$D_1\equiv D$} and~\mbox{$E_1\equiv E$}.  
One can see that for uniaxial anisotropy, the eigenstates are conveniently labeled with~$M_n$ and they are degenerate when having equal $|M_n|$. However, for non-vanishing transverse anisotropy (\mbox{$E\neq0$}) the degeneracy of charged states,
$
	\ket{\Chi_1}
	\in
	\big\{
	\mbox{$
	\ket{\chi^0_1}
	\equiv
	\ket{0}$},
	\ket{\chi_1^\pm}
	\equiv
	\big(\ket{1}\pm\ket{-1}\big)/\sqrt{2}
	\big\}
$,	
is lifted. 

\subsection{Impact of molecular vibrations}
\label{sec:model_vib}

Importantly, a molecule embedded in a junction generally supports different vibrational modes. These vibrational modes are approximated as independent harmonic oscillators~\cite{Mahan_book} with angular frequencies~$\omega_q$, 
\begin{equation}\label{eq:H_vib}
	\Hvib
	=
	\sum\limits_{q=1}^{Q}
	\hbar
	\omega_q
	\opb_q^\dagger\opb_q^{} 	
	+
	\Hchvib
	+
	\Hspinvib	,
\end{equation}
and they can in principle couple \emph{both} to the charge ($\Hchvib$) and spin ($\Hspinvib$) degrees of freedom of the molecule.
The operator~$\opb_q^\dagger$\,($\opb_q^{}$) denotes the creation (annihilation) operator for the $q$th quantized vibrational mode, referred commonly to as a \emph{vibron}.  We here assume the total number of vibrational modes to be $Q$.
In the absence of the coupling terms, $\Hspinvib$ and $\Hchvib$, the vibronic contribution, $\ket{\vartheta}$, to the molecular eigenstates is given by 
\mbox{$
	\ketv{\vartheta}
	\equiv
	\ketv{\nv^1,\ldots,\nv^Q}
$}
with eigenenergies
\mbox{$
	\en_\vartheta
	=
	\sum_{q=1}^Q
	\hbar\omega_q\nv^q
$,}
where~$\nv^q$ is the occupation number of the $q$th vibrational mode.

The coupling of these vibrations to the electronic charge has been extensively studied~\cite{Mahan_book,Mitra2004,Koch2005,Koch2006Nov}, and is captured by the Hamiltonian 
\begin{equation}\label{eq:H_ch-vib}
	\Hchvib
	=
	\sum\limits_{q=1}^{Q}
	\lq
	\hbar\omega_q
	\big(\opb_q^\dagger+\opb_q^{}\big)
	\opn
	,
\end{equation}
with the dimensionless coupling strength~$\lq$.

However, in a molecule in which deformations (for example, due to the embedding into the junction) influence its magnetic anisotropy~\cite{May2011,Ruiz2012}, small oscillations around the equilibrium position, are expected to lead to interactions between molecular vibrations and the spin as well~\cite{Kokado2010Spin,Ruiz2012}. This is represented by the third term of the Hamiltonian~(\ref{eq:H_vib}), 
\begin{equation}\label{eq:H_spin-vib}
	\Hspinvib
 	=
 	\sum\limits_{n=0,1}
 	\sum\limits_{q=1}^{Q}
 	\hbar \omega_q
 	\,
 	\opSVnq
 	\big(\opb_q^\dagger + \opb_q^{}\big)
 	.
\end{equation}
Here, the operator~$\opSVnq$ reads as
\begin{equation}\label{eq:opSV_def}
	\opSVnq
 	=
 	\Luqn
 	\big(\opS_n^z\big)^{\!2}
 	+
	\Ltqn
 	\Big[  
 	\big(\opS_n^x\big)^2 
 	-
 	\big(\opS_n^y\big)^2
 	\Big]
 	,
\end{equation}
and~the dimensionless parameters~$\Luqn$ and~$\Ltqn$ stand for the coupling strength of vibrations to the uniaxial and transverse components of the molecular spin, respectively. 
%
%
In the following discussion,~$\omega_q$, $\lq$, $\Luqn$ and~$\Ltqn$, as well as $D_n$ and $E_n$ are treated as tunable, continuous parameters. A possibility to address the strength of the magnetic anisotropy in a molecule is \via stretching in a break junction setup~\cite{May2011,Ruiz2012}. Note that in such a case also the vibration frequency and the strength of the coupling to the charge are tunable \via the junction properties~\cite{Xiang2016,Kim2011,Bruot2011}.

Finally, it should be mentioned that in general the operator~$\opSVnq$ can take a more complex form, determined by the symmetry properties of the molecular spin and vibrations depending on how the molecule is embedded in the junction. In other words, it is conditioned by how the coupling to the electrodes of the junction and the molecular vibrations affect the ligand field, generating thus additional contributions to the magnetic anisotropy of the molecule~\cite{Garanin1997,Ganzhorn2013}.

\subsection{Tunnel coupling to electrodes}
\label{sec:model_tun}

The embedding of the molecule into an electronic junction enables electron tunneling processes between junction and molecule, which thereby change the charge- and spin-state of the molecule. 
Within the model under consideration, the electrodes of the tunnel junction are represented as two reservoirs of non-interacting electrons as described by the first term of the Hamiltonian
\begin{equation}\label{eq:H_tun}
	\Hjun
	= 
	\sum_{rk\sigma}
	\varepsilon_{k\sigma}^r
	\opa_{k\sigma}^{r\dagger}
	\opa_{k\sigma}^{r\phantom{\dagger}} 
	+
	\sum_{rlk\sigma}\left(
	t_l^r
	\opd_{l\sigma}^\dagger
	\opa_{k\sigma}^{r\phantom{\dagger}}
	+
	\text{H.c.}\right)
	.
\end{equation}
The operator~$\opa_{k\sigma}^{r\dagger}$\,($\opa_{k\sigma}^{r\phantom{\dagger}}$) is responsible for  creation (an\-ni\-hilation) of an electron with energy~$\varepsilon_{k\sigma}^r$ in~drain (\mbox{$r=\text{D}$}) and source (\mbox{$r=\text{S}$}) electrodes, with~$k$ and~$\sigma$ denoting the orbital and spin quantum numbers, respectively. 
Furthermore, the electronic occupation of the electrodes is governed by Fermi functions, \mbox{$f_r(\epsilon)\!=\!\big\{1+\exp[(\epsilon-\mu_r)/(\kB T)]\big\}^{-1}$}, with temperature~$T$ and a possible bias (transport) voltage~$\Vb$ given by the difference of electrochemical potentials of the electrodes, \mbox{$\Vb=(\mu_\text{S}-\mu_\text{D})/e$}.

Next, electron tunneling processes between electrodes and the molecule are included in the second term of Eq.~(\ref{eq:H_tun}), where~$t_l^r$ is the (spin-independent) tunneling matrix element between the~$l$th molecular orbital and the $r$th electrode.
A very convenient basis for studying transport of electrons is the basis of molecular states~\mbox{$\big\{\ket{\Chi_n}\!\otimes\!\ketv{\vartheta}\big\}$}. The tunneling Hamiltonian [that is, the second term of Eq.~(\ref{eq:H_tun})] expanded in this basis takes the form~\cite{Misiorny2015}
\begin{equation}\label{eq:H_tun_expand}
	\Htun
	=
	\sum_{rk\sigma}
	\sum_{\Chi_0\Chi_1\vartheta}
	\!\!
	\mathbb{T}_r
	\tunT_{\Chi_1 \Chi_0}
	\ket{\Chi_1}\bra{\Chi_0}
	\otimes
	{\ket{\vartheta}\bra{\vartheta}}
	\,
	\opa_{k\sigma}^{r\phantom{\dagger}}
	+
	\text{H.c.}
\end{equation}
In the equation above, we split the tunneling amplitude into two factors: one quantifying the orbital overlap of the molecular states ($\mathbb{T}_r$), and the other imposing spin selection rules for transitions between molecular states ($\tunT_{\Chi_1 \Chi_0}$).
The former is given by 
\mbox{$
	\mathbb{T}_r 
	= 
	\sum_l 
	t^r_l   
	\bra{S_1}| \opd_{l}^\dagger| \ket{S_0}
$},
with \mbox{$\bra{S_1}|\opd_{l}^\dagger|\ket{S_0}$} denoting the so-called \emph{reduced matrix element}, which occurs here due to application of the Wigner-Eckart theorem~\cite{Sakurai_book}. 
The explicit form of the latter is
\begin{equation}\label{eq:tunT_def}
	\tunT_{\Chi_1 \Chi_0}
	= 
	\sum_{M_0 M_1} 
	\!\!
	\big(\coefC_{M_1}^{\Chi_1}\big)^{\!\ast}
	\coefC_{\Chi_0}^{M_0}
	\,
	\big\langle S_0,M_0;\tfrac{1}{2},\sigma\big|S_1, M_1\big\rangle
\end{equation}
with \mbox{$\big\langle S_0,M_0;\tfrac{1}{2},\sigma\big|S_1, M_1\big\rangle$} standing for the  Clebsch-Gordon coefficient. 
Moreover, $\mathbb{T}_r$ is treated here as a free parameter. It enters the spin-de\-pend\-ent broadening~$\Gamma_\sigma^r$ of molecular levels, \mbox{$\Gamma_\sigma^r=2\pi\nu_\sigma^r|\mathbb{T}_r|^2$}, which arises as a result of tunneling of electrons between a molecule and the $r$th electrode. The coefficient~$\nu_\sigma^r$ stands for the spin-resolved density of states (DOS) in the $r$th electrode in a flat-band approximation [namely, the DOS is assumed to be energy-independent, $\nu_\sigma^r(\varepsilon)\approx\nu_\sigma^r$].

In the following, we allow the electrodes to be spin-polarized. 
Note that only a collinear relative orientation of the spin moments in the electrodes \mbox{---that} is, the \emph{parallel} and \emph{antiparallel} magnetic configuration, as shown in Fig.~\ref{fig1}(a)--- is considered, and we take these spin moments also to be collinear with the principle ($z$)~axis of the molecule.
To quantify the~magnetic properties of the electrodes we introduce the spin-polarization coefficient~$P_r$ defined in terms of the DOS of spin-majority (\mbox{-mi}\-nor\-i\-ty) electrons, \mbox{$\nu_{+(-)}^r$}, as
\mbox{$
	P_r
	=
	(\nu_+^r-\nu_-^r)
	/
	(\nu_+^r+\nu_-^r)
$}.
For equal spin-polarizations of the two electrodes (\mbox{$P_\text{S}=P_\text{D}\equiv P$}) and for symmetric tunnel-coupling (\mbox{$\mathbb{T}_\text{S}=\mathbb{T}_\text{D}\equiv\mathbb{T}$}), assumed henceforth, we can parametrize~$\Gamma_\sigma^r$ in terms of the spin-polarization coefficient~$P$ and the total broadening~\mbox{$\Gamma\equiv\Gamma^r=\Gamma_\uparrow^r+\Gamma_\downarrow^r$} as follows:
\mbox{$\Gamma_{\uparrow(\downarrow)}^\text{S}=\Gamma_{\uparrow(\downarrow)}^\text{D}=(\Gamma/2)(1\pm P)$} for the parallel magnetic configuration, and \mbox{$\Gamma_{\uparrow(\downarrow)}^\text{S}=\Gamma_{\downarrow(\uparrow)}^\text{D}=(\Gamma/2)(1\pm P)$} for the antiparallel one.
%

\section{Effective Hamiltonians}
\label{sec:states_and_effHams}

%
Due to the coupling between vibrations and the molecule's charge  and spin degrees of freedom, see Eqs.~(\ref{eq:H_ch-vib})-(\ref{eq:H_spin-vib}), the molecular states~\mbox{$\big\{\ket{\Chi_n}\!\otimes\!\ketv{\vartheta}\big\}$} are not eigenstates of the Hamiltonian $\Hmol+\Hvib$ any longer.
The purpose of this section is to eliminate the charge-vibron and spin-vibron couplings from the Hamiltonian~$\Hmol+\Hvib$ by application of appropriate canonical transformations, 
\begin{equation}\label{eq:Canom_Transform}
	\big(
	\Hmol+\Hvib
	\big)^{\!\prime}
	=
	\text{e}^{\opAsv}
	\text{e}^{\opAcv}
	\big(
	\Hmol+\Hvib
	\big)
	\text{e}^{-\opAcv}
	\text{e}^{-\opAsv}
	.
\end{equation}
The scope of this transformation is that the new effective Hamiltonian~\mbox{$\big(\Hmol+\Hvib\big)^{\!\prime}$} \mbox{---with} renormalized \mbox{parameters---} becomes diagonal in the basis~\mbox{$\big\{\ket{\Chi_n}\otimes\ketv{\vartheta}\big\}$}. Particularly, the transformation kernels~$\opAcv$ and~$\opAsv$ allow for elimination of the charge-vibron~($\Hchvib$) and spin-vibron~($\Hspinvib$) interactions, respectively.
%

\subsection{Charge-vibron coupling in the absence of spin-vibron coupling}
\label{sec:chargevibron}

The former kernel, first introduced by Lang and Firsov~\cite{Lang1963}, is known to have the form
\mbox{$
	\opAcv
	=
	\sum_{q=1}^Q
	\lq
	\big(\opb_q^\dagger-\opb_q^{\phantom{\dagger}}\big)\opn
$,}
and it has proven very useful for studying the Franck-Condon phenomena in  transport through single-molecule devices~\cite{Mitra2004,Koch2005,Koch2006Nov,McCaskey2015}. The Lang-Firsov transformation decouples the charge and vibronic operators, leading at the same time to an energy shift of the  charged state, 
\mbox{$
	\en(\Vg)
	\mapsto
	\en(\Vg)
	-
	\sum_{q=1}^Q
	\hbar\omega_q\lq^2
$.}
Importantly, at the same time also the tunneling Hamiltonian~(\ref{eq:H_tun_expand}) gets modified
\begin{multline}\label{eq:H_tun_expand_LF}
	\text{e}^{\opAcv}
	\Htun
	\text{e}^{-\opAcv}
	=
	\mathbb{T}
	\sum_{rk\sigma}
	\sum_{\Chi_0\Chi_1}
	\sum_{\vartheta\vartheta'}
	\tunT_{\Chi_1 \Chi_0}
	\FCcoef_{\vartheta'\vartheta}^{}
\\
	\ket{\Chi_1}\bra{\Chi_0}
	\otimes
	{\ket{\vartheta'}\bra{\vartheta}}
	\,
	\opa_{k\sigma}^{r\phantom{\dagger}}
	+
	\text{H.c.}
\end{multline}
Note that in this transformed tunneling Hamiltonian the number of vibrational excitations is not conserved anymore.
The new coefficient~$\FCcoef_{\vartheta'\vartheta}$ is the so-called Franck-Condon matrix element~\cite{Koch2006Nov,Seldenthuis2008,Cuevas-Scheer_book}, 
\begin{equation}\label{eq:FCcoef_def}
	\FCcoef_{\vartheta'\vartheta}
	=
	\brav{\vartheta'}
	\exp\Big[
	\sum_{q=1}^Q
	\lq\big(\opb_q^\dagger-\opb_q^{}\big)
	\Big]
	\ketv{\vartheta}
	.
\end{equation}
%

\subsection{Spin-vibron coupling}
\label{sec:spinvibron}

In the presence of spin-vibron interaction, Eq.~(\ref{eq:H_spin-vib}), the Lang-Firsov transformation generates an additional term in the molecular Hamiltonian,
\begin{equation}
\hspace*{-4pt}
	\text{e}^{\opAcv}
	\Hspinvib
	\text{e}^{-\opAcv}
	\!
	=
	\Hspinvib
	-
	2
	\!
 	\sum\limits_{q=1}^{Q}
 	\lq
 	\hbar \omega_q
 	\opSV_{\!1q}
 	\opn
 	.
 	\!
\end{equation}
Noticeably, this term does not couple spin and vibrational degrees of freedom of the molecule, but represents a correction to the magnetic anisotropy of the molecule in the charged state.

In a next step, we derive the kernel~$\opAsv$ of the canonical transformation~(\ref{eq:Canom_Transform}), which can remove the spin-vibron interaction leading to an effective molecular Hamiltonian with renormalized magnetic-anisotropy parameters. The following discussion is divided into two parts: first, we consider molecules with uniaxial anisotropy, only, (that is, with \mbox{$E_n=0$} and \mbox{$\Ltqn=0$}), and second, we cover the more general case of molecules exhibiting both uniaxial and transverse anisotropy.

\subsubsection{Molecules with purely uniaxial magnetic anisotropy}
\label{sec:EffH_uni}

For this first case, we set \mbox{$E_n=0$} in Eq.~(\ref{eq:H_spin}) and  \mbox{$\Ltqn=0$} in Eq.~(\ref{eq:opSV_def}).
In order to derive the transformation kernel~$\opAsv$, we apply the procedure described in Ref.~\cite{Wagner_book}, projecting the spin-vibron interaction term~$\Hspinvib$, Eq.~(\ref{eq:H_spin-vib}), on the states  \mbox{$\big\{\ket{M_n}\otimes\ketv{\vartheta}\big\}$}, which are the eigenstates of the~Hamiltonian 
\mbox{$
	\Ham_0 
	= 
	\text{e}^{\opAcv}
	\big(
	\Hmol +\sum
	\hbar
	\omega_q
	\opb_q^\dagger\opb_q^{} 
+ \Hchvib
	\big)
	\text{e}^{-\opAcv}
$.}
With this we find
\begin{equation}\label{eq:opAs_1_def}
	\opAsv
	=
	\sum_{n=0,1}
	\sum\limits_{q=1}^Q
	\Luqn 
	\big(\opS_n^z\big)^2 
	\big(\opb_q^\dagger - \opb_q^{}\big)
	.
\end{equation}
This expression agrees with that used by Ruiz-Tijerina~\etal~\cite{Ruiz2012}, who studied the effect of magnetic anisotropy dynamically induced by mechanical stretching of a molecule on transport in the Kondo regime.
Next, inserting the operator~(\ref{eq:opAs_1_def}) into Eq.~(\ref{eq:Canom_Transform}), we obtain the effective (renormalized) Hamiltonian of the molecule with vibrations
%
$	\Hg^\prime
	+
	\Hspin^\prime
	+
	\sum
	\hbar
	\omega_q
	\opb_q^\dagger\opb_q^{} 
	.
$
%
Here, the charge part of the molecular Hamiltonian is given by
\begin{equation}\label{eq:H_ch_1}
	\Hg^\prime
	=
	\Big[
	\en(\Vg)
	-
	\sum_{q=1}^Q
	\hbar\omega_q\lq^2
	\Big]\opn
	,
\end{equation}
with the energy shift caused by the charge-vibron interaction, as mentioned above. 
Importantly, the spin-vibron coupling is eliminated at the expense of modifying the \emph{magnetic} properties of the molecule, and the spin term~$\Hspin^\prime$ is written as
\begin{equation}\label{eq:H_spin_1}
	\Hspin^\prime
	= 
	-
	\sum_{n=0,1}
	\!
	\Big[
	\big(D_n+\Dnt\big)
	\big(\opS_n^z\big)^{\!2}
	+
	\Dnf
	\big(\opS_n^z\big)^{\!4}
	\Big]
	.
\end{equation}
The anisotropy is affected in two ways: First, the uniaxial anisotropy constant~$D_n$ in Eq.~(\ref{eq:H_spin}) is renormalized as \mbox{$D_n\mapsto D_n+\Dnt$}, with 
\begin{equation}\label{eq:Dnt_def}
	\Dnt
	=
	2\delta_{n1}\sum_{q=1}^Q
	\lq
	\Lambda_{1q}^\text{u}	
	\hbar\omega_q
	.
\end{equation}
Second, a new component representing a fourth-order-in-spin contribution to the uniaxial magnetic anisotropy \big[\mbox{$\propto(\opS_n^z)^4$}\big] appears in Eq.~(\ref{eq:H_spin_1}),  with the aniso\-tropy constant~$\Dnf$ taking the form
\begin{equation}\label{eq:Dnf_def}
	\Dnf
	=
	\sum_{q=1}^Q
	\big(\Luqn\big)^{\!2}
	\hbar\omega_q
	.
\end{equation}
The result of Eqs.~(\ref{eq:H_ch_1})-(\ref{eq:Dnf_def}) is an effective molecular Hamiltonian, which is diagonal in the basis of product states~\mbox{$\big\{\ket{M_n}\otimes\ketv{\vartheta}\big\}$}. Note that the transformation with the operator $\opAsv$ does not further affect the tunneling Hamiltonian given in Eq.~(\ref{eq:H_tun_expand_LF}).%
\footnote{%
Note that this comes as a consequence of the present approximation that the effective molecular spin in Eq.~(\ref{eq:H_spin}) arises as a result of stabilization of a large atomic spin in the presence of the crystal/ligand field. However, in the case when the effective spin can be derived from a microscopic model of interacting electrons in different molecular orbitals, one generally expects that the transformation with the operator~$\opAsv$ can lead to occurrence of new effective tunneling matrix elements that depend on the magnetic states of the molecule, as shown in Ref.~\cite{Ruiz2012}.%
}

\newpage

\subsubsection{Molecules with uniaxial\\ and transverse magnetic anisotropy}
\label{sec:EffH_uni_and_trans}

The situation becomes more complicated for a molecule with an additional non-vanishing transverse component of magnetic anisotropy~(\mbox{$E_n\neq0$}).
In general, there exists no generic canonical transformation that would allow for \emph{exact} elimination of the spin-vibron coupling. 
The reason is that Hamiltonians~$\Hspinvib$ and~$\Ham_0$ do not share the same basis of eigenstates, that is,~\mbox{$\big[\Hspinvib,\Ham_0\big]\neq0$}, and, hence, the full molecular Hamiltonian~\mbox{$\Hmol+\Hvib$} [see Eq.~(\ref{eq:H_mol}) and Eq.~(\ref{eq:H_vib})] cannot be diagonal with respect to both~$ \hat{\mathcal{H}}_0 $ and~$ \Hspinvib $ simultaneously.
Nonetheless, there are two particular cases for which commutation of~$\Hspinvib$ and~$\Ham_0$ can be restored so that they can be diagonalized in the basis~\mbox{$\big\{\ket{\Chi_n}\otimes\ketv{\vartheta}\big\}$}:
the first one resorts to a specific constraint of parameters (namely, if
\mbox{$  
	D_n\Ltqn 
	= 
	-{E_n}\Luqn
$}),
while the second one exploits the fact that ---independently of the anisotropy parameters--- ~\mbox{$\big[\Hspinvib,\Ham_0\big]=0$} for a molecular spin~\mbox{$S_n\leqslant1$}. The key advantage in the latter case is that, though not applicable to large-spin molecules (\ie, with \mbox{$S_n>1$}), this solution does not involve any additional restrictions regarding the properties of the molecule.

In either of these cases, the same method as in Sec.~\ref{sec:EffH_uni} can be used and we obtain
\begin{equation}\label{eq:opAs_2_def}
	\opAsv
 	=
	\sum_{n=0,1}
	\sum\limits_{q=1}^Q
 	\opSVnq
 	\big(\opb_q^\dagger - \opb_q^{})
 	.
\end{equation}
The effective giant-spin Hamiltonian now reads as
\begin{align}\label{eq:H_spin_2}
	\Hspin^\prime 
	= 
	\sum_{n=0,1}
	\!
	\Big[
	&
	\!
	-\big(D_n+\Dnt\big)
	\big(\opS_n^z\big)^{\!2}
	-\Dnf
	\big(\opS_n^z\big)^{\!4}
\nonumber\\[-7pt]
	&\!
	+
	\big(E_n+\Ent\big)
	\Big[
	\big(\opS_n^x\big)^{\!2}
	-
	\big(\opS_n^y\big)^{\!2}
	\Big]
\nonumber\\
	&\!	
	+
	\Enf
	\Big[
	\big(\opS_n^x\big)^{\!2}
	-
	\big(\opS_n^y\big)^{\!2}
	\Big]^2
\nonumber\\
	&\!
	+
	\Cnf
	\Big\{
	\big(\opS_n^z\big)^{\!2}
	,
	\big(\opS_n^x\big)^{\!2}
	-
	\big(\opS_n^y\big)^{\!2}
	\Big\}
	\Big]
	,
\end{align}
where~$\{\bullet,\bullet\}$ in the last line denotes the anticommutator. The corrections~$\Dnt$ and~$\Dnf$ are given by Eq.~(\ref{eq:Dnt_def}) and Eq.~(\ref{eq:Dnf_def}), respectively, while the remaining corrections are found to be
\begin{gather}\label{eq:Ent_def}
	\Ent
	=
	-
	2\delta_{n1}\sum_{q=1}^Q
	\lq
	\Lambda_{1q}^\text{t}	
	\hbar\omega_q
	,
\\\label{eq:Enf_def}
	\Enf
	=
	-
	\sum_{q=1}^Q
	\big(\Ltqn\big)^{\!2}
	\hbar\omega_q
	,
\\\label{eq:Cnf_def}
	\Cnf
	=
	-
	\sum_{q=1}^Q
	\Luqn\Ltqn
	\hbar\omega_q
	.
\end{gather}
It means that in addition to the renormalization of the \emph{strength} of the uniaxial and transverse anisotropy, in general an additional \emph{type} of anisotropy is introduced by the combined uniaxial and transverse spin-vibron coupling.

Consequently, the coupling of vibrations to the charge and spin of a molecule modifies its energy spectrum in various ways.
In the remainder of this paper, we consider these effects for different example molecules and study both the explicit impact on the energy spectra, Sec.~\ref{sec:spectrum}, as well as the resulting features expected to appear in the tunneling current through these molecules when embedded into a junction, Sec.~\ref{sec:transport}. 
%

\section{Impact on spectral properties}
\label{sec:spectrum}

The first, obvious impact of vibrations on the spectrum of a molecule manifests as a repetition of the magnetic spectrum of the static molecule at energies corresponding to multiples of the energies~$\hbar \omega_q$ of the vibrational modes~\mbox{$q=1,\dots,Q$}. 
This indeed plays a role in transport properties, as will be studied in detail in Sec.~\ref{sec:transport}, where transitions between states with different vibronic occupations occur. 
In the present section, we concentrate on the \textit{nontrivial} impact of vibrations \mbox{---resulting} from the \emph{coupling} between vibrations and the charge and spin of the {molecule---} on the magnetic component of the molecular spectrum. 
Since this part of the spectrum becomes modified identically in all vibrational states, below we simply focus on the vibrational ground state (with \mbox{$\nv^q=0$} for all~$q$).

\subsection{Interplay of magnetic anisotropy and vibrations}
\label{sec:E_D_spectrum}

In this subsection, employing the example molecule introduced in Sec.~\ref{sec:model_mol} with the ``static'' energy spectrum shown in Fig.~\ref{fig1}(b), we will illustrate how vibrations affect the magnetic spectrum of a molecule. To begin with, recall that in the neutral state  this model molecule is characterized by a spin~\mbox{$S_0=1/2$}, corresponding to a spin doublet, \mbox{$\ket{\chi_0^\pm}\equiv\ket{\pm1/2}$}.
From Eqs.~(\ref{eq:H_ch_1})-(\ref{eq:Dnf_def}) and Eqs.~(\ref{eq:H_spin_2})-(\ref{eq:Cnf_def}), one finds that the spin-vibron interaction only results in an energy shift \mbox{$\Delta_0=-\Dnf/16$}. 
The situation is different in the charged state, characterized by a spin~\mbox{$S_1=1$}, in which the magnetic state of a molecule is the spin triplet:
\mbox{
$	
	\ket{\chi_1^0}
	=
	\ket{0}
$%
}
and
$
	\ket{\chi_1^\pm}
	=
	\big(
	\mbox{$\ket{1}$}
	\pm
	\mbox{$\ket{-1}$}
	\big)
	/
	\sqrt{2}
	.
$
In such a case,~we can simplify the effective spin Hamiltonian~$\Hspin^\prime$, Eq.~(\ref{eq:H_spin_2}),  to
\begin{equation}\label{eq:H_spin_eff}
	\Ham_{\text{spin},n=1}^\prime
	=
	-\Deff
	\big(\opS_1^z\big)^{\!2}
	+
	\Eeff
	\Big[
	\big(\opS_1^x\big)^{\!2}
	-
	\big(\opS_1^y\big)^{\!2}
	\Big]
	,
\end{equation}
where $\Deff=D+\Delta D$ and $\Eeff=E+\Delta E$ with
\begin{gather}\label{eq:DeltaD_def}
	\Delta D
	=
	\delta D_1^{(2)}
	+
	\delta D_1^{(4)}
	-
	\delta E_1^{(4)}
	,
\\\label{eq:DeltaE_def}
	\Delta E
	=
	\delta E_1^{(2)}
	+
	2\delta C_1^{(4)}
	.
\end{gather}
We remind that due to the capacitive coupling of the molecule to a gate electrode, the relative position of the neutral doublet and the charged triplet can be continuously adjusted by application of the gate voltage~$\Vg$. For instance, it allows for compensating the shift~$\Delta_0$. This shift will therefore be omitted from now on.

To further discuss the impact on the spectrum, we assume for simplicity that only one vibrational mode of energy~$\hbar\omega$ is involved in the coupling (we hence omit the vibrational mode index `$q$'). In this example,  we also take the anisotropy constants~$D$ and~$E$, as well as all coupling parameters to be positive; the case of \mbox{$D<0$} is analyzed in Sec.~\ref{sec:barrier_flip}.
The corrections to the magnetic anisotropy, Eqs.~(\ref{eq:DeltaD_def})-(\ref{eq:DeltaE_def}),  take then the explicit form,
\begin{gather}\label{eq:dD_magnitude}
	\frac{\Delta D}{\lambda\Lu_1\hbar\omega}
	=
	2
	+
	\frac{\Lu_1}{\lambda}
	\big(1+\zeta^2\big)
	,
\\\label{eq:dE_magnitude}
	\frac{\Delta E}{\lambda\Lu_1\hbar\omega}
	=
	-
	2\zeta
	\bigg[
	1
	+
	\frac{\Lu_1}{\lambda}
	\bigg]
	,
\end{gather}
where we introduce the coefficient~\mbox{$\zeta=\Lt_1/\Lu_1$}. 
Let us make an estimate of the relevance of these corrections with respect to the original anisotropy parameters~$D$ and~$E$. Both corrections depend linearly on the charge-vibron coupling strength~$\lambda$ and  the energy of the vibrational mode~$\hbar\omega$.  
In general, one expects that the charge-vibron interaction dominates over the spin-vibron coupling, that is, \mbox{$\Lu_1/\lambda\ll1$}. In this case, we can approximate \mbox{$\Delta D\approx2\lambda\Lu_1\hbar\omega$} and \mbox{$\Delta E\approx-2\lambda\Lt_1\hbar\omega$}.
Since the energy of the vibrational mode~$\hbar\omega$ can be significantly larger than the magnetic anisotropy~$D$, \mbox{$\hbar\omega\gg D$}~\cite{Burzur2014,McCaskey2015}, we conclude that even if the charge- and spin-vibron couplings are not particularly strong (\mbox{$\lambda\lesssim1$} and \mbox{$\Lu_1/\lambda\ll1$}), the shift~$\Delta D$  can still achieve appreciable values compared to $D$ (and equivalently for~$\Delta E$ and $E$).

\begin{figure}[t!]
	\includegraphics[width=0.99\columnwidth]{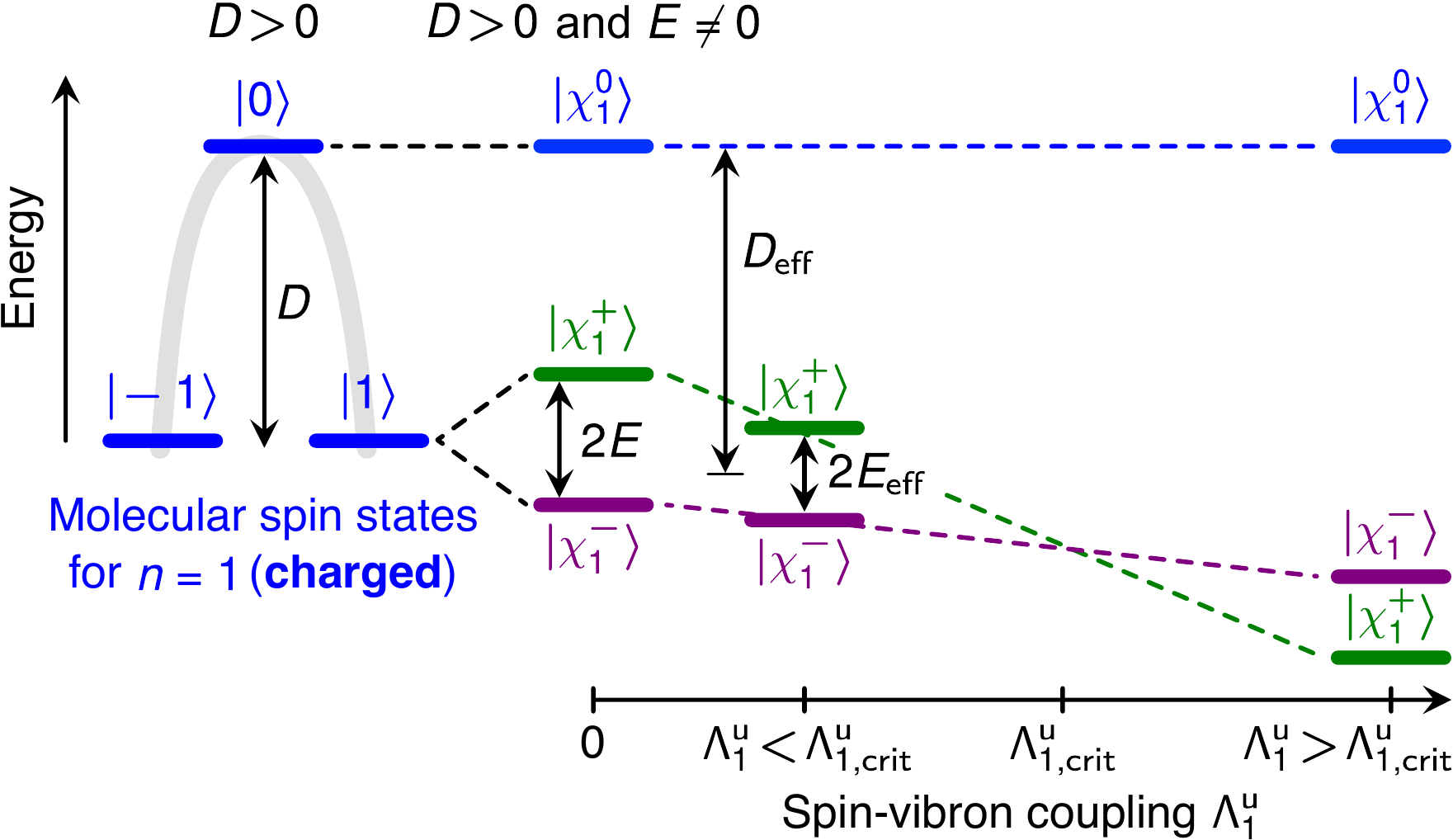}
	\caption{%
	Effect of the charge- and spin-vibron coupling for~\mbox{$\zeta\equiv\Lt_1/\Lu_1<1$}, fixed $\lambda$ and a single vibrational mode of energy~$\hbar\omega$ illustrated for continuously changing values of the spin-vibron coupling~$\Lu_1$.
	At the \emph{critical} spin-vibron coupling~$\Lucrit$ the effective transverse magnetic anisotropy becomes suppressed, that is, the states~$\ket{\chi_1^+}$ and $\ket{\chi_1^-}$ are degenerate, see Eq.~(\ref{eq:Lu_crit}).
	For \mbox{$\Lu_1<\Lucrit$}, $D$ is effectively increased while $E$ is effectively reduced, and  for \mbox{$\Lu_1>\Lucrit$}, the energies of the two states are inverted.
	Further details can be found in Sec.~\ref{sec:E_D_spectrum}.
		}
	\label{fig2}
\end{figure}

In Fig.~\ref{fig2}, we schematically show how the spin-vibron coupling affects the energy of the spin states,
\mbox{$
	\en_{\chi_1^\pm}
	=
	-\Deff\pm\Eeff
$}, taking \mbox{$\en_{\chi_1^0}=0$} as reference energy. 
Specifically, we tune the uniaxial component of the spin-vibron  coupling~$\Lu_1$ here, while for simplicity fixing the vibration energy $\hbar\omega$, the charge-vibron coupling strength~$\lambda$, as well as the relation between~$\Lu_1$ and~$\Lt_1$ given by~$\zeta$, focusing on a value \mbox{$\zeta<1$}. Nonetheless,  we recall that due to the deformation of a molecule, all parameters~$\omega$, $\lambda$, $\Lu_1$ and~$\Lt_1$ can in principle change.

First of all, it can be seen that the states~$\ket{\chi_1^-}$ and~$\ket{\chi_1^+}$ respond differently to changing~$\Lu_1$. 
Since~$\Delta D$ is positive [see Eq.~(\ref{eq:dD_magnitude})], whereas~$\Delta E$ is negative [see Eq.~(\ref{eq:dE_magnitude})], their impact on the two states is also not equally strong: While for~$\ket{\chi_1^+}$ the effect of these two corrections is additive, \mbox{$-\Delta D-|\Delta E|$}, the effect on~$\ket{\chi_1^-}$ is reduced, namely, it is 
\mbox{$-\Delta D+|\Delta E|$}.\footnote{%
In particular, if $\zeta$ was increased such that~$\zeta\approx1$, one would find $\Delta D\approx|\Delta E|$ and the effect of the spin-vibron coupling on $\ket{\chi_1^-}$ would be completely suppressed.
}
A further result of this dissimilar behavior of~$\ket{\chi_1^+}$ and~$\ket{\chi_1^-}$ is that their energies can, in general, even be inverted with increasing~$\Lu_1$. The crossover between these two situations happens at a critical value~$\Lucrit$, namely at
\begin{equation}\label{eq:Lu_crit}
	\Lucrit
	=
	\sqrt{
	\bigg(\!\frac{\lambda}{2}\!\bigg)^{\!\!2}
	+
	\frac{E}{2\zeta\hbar\omega}
	}
	-
	\frac{\lambda}{2}
	,	
\end{equation}
where the degeneracy of the states~$\ket{\chi_1^+}$ and~$\ket{\chi_1^-}$ is restored (\mbox{$\en_{\chi_1^+}=\en_{\chi_1^-}$}). Gaining control over the spin-vibron coupling is therefore extremely advantageous, because it would enable enhancing the overall anisotropy (important for information storage) and at the same time it could reduce, or even fully cancel, the energy splitting between the lower lying states.

The value of $\zeta$ determines the slope of the energy of the state $\ket{\chi_1^-}$ as a function of the spin-vibron coupling (shown in Fig.~\ref{fig2} for a negative slope at $\zeta<1$). Thus, if a molecule is characterized by~\mbox{$\zeta>1$} (that is, when vibrations couple stronger to the transverse component of the molecular spin) and by vibrational modes of sufficiently large energies, it is actually possible that ---due to a large \textit{positive} slope--- the energy of~$\ket{\chi_1^-}$ can become larger than that of~$\ket{\chi_1^0}$. 
More generally speaking, the value of $\zeta$ influences the energy at which the states $\ket{\chi_1^-}$ and $\ket{\chi_1^+}$ cross as well as the critical spin-vibron coupling at which the crossing occurs [see also Eq.~(\ref{eq:Lu_crit}) for the dependence of the critical coupling on $\zeta$]. For this reason, in Sec.~\ref{sec:Effect_of_zeta}, we will also discuss how the value of $\zeta$ affects the transport characteristics of the system.

\subsection{Magnetic spectrum reversal}
\label{sec:barrier_flip}

In general, the sign of corrections to the magnetic anisotropy due to spin-vibron coupling depends on whether the relevant coupling parameters~$\Luqn$ and~$\Ltqn$ have the same or opposite signs with respect to the bare anisotropy parameters~$D_n$ and~$E_n$, see Eqs.~(\ref{eq:Dnt_def})-(\ref{eq:Dnf_def}) and Eqs.~(\ref{eq:Ent_def})-(\ref{eq:Cnf_def}) in Sec.~\ref{sec:states_and_effHams}. 
In the previous subsection, we have fixed all these parameters to be \emph{positive}. In consequence,  we have concluded that while the correction~$\Delta D$ to the uniaxial component of magnetic anisotropy is expected to be positive [see Eq.~(\ref{eq:dD_magnitude})], the correction~$\Delta E$ to the transverse component~$\Eeff$ is negative [see Eq.~(\ref{eq:dE_magnitude})]. The latter can result in quenching the transverse anisotropy for some particular values of the spin-vibron couplings.
One should, however, notice that molecules can also be characterized by one or both \emph{negative} bare anisotropy parameters. Interestingly, in such a case we predict that the coupling of charge and spin of a molecule to its vibrations can lead to a substantial qualitative change of the magnetic spectrum. This effect may play a key role especially for a large-spin molecule (that is, with~\mbox{$S_0,S_1>1$} and~\mbox{$|S_1-S_0|=1/2$}) and in the absence of transverse magnetic anisotropy (\mbox{$E_0=E_1=0$}), where it can be observed in transport measurements as the onset of a pronounced spin blockade, as we will show in Sec.~\ref{sec:transport_blockade}.

\begin{figure}[t]
	\includegraphics[scale=1]{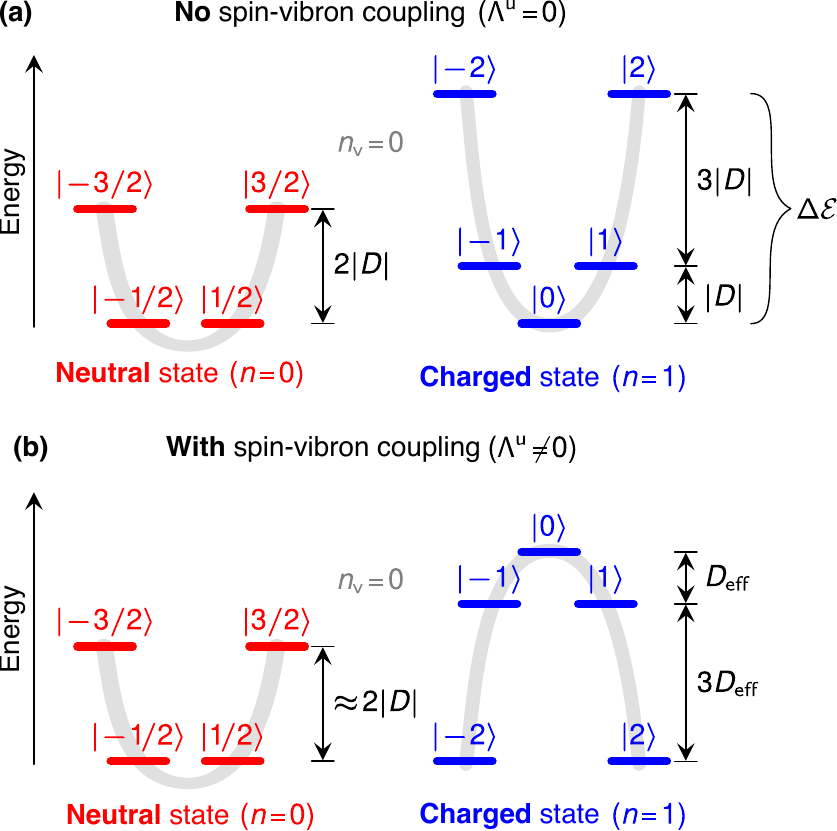}
	\caption{
	Energy spectrum of an exemplary molecule with \mbox{$S_0=3/2$} and \mbox{$S_1=2$} with \mbox{$D\equiv D_0=D_1<0$} and \mbox{$E_n=0$}. 
	(a)~No spin-vibron coupling (\mbox{$\Lu=0$}). (b) Modification of the molecular spectrum due to \mbox{$\Lu\neq0$}, with \mbox{$\Lu\equiv\Lu_0=\Lu_1$} and \mbox{$\Deff\approx D+2\lambda\Lu\hbar\omega$}.
	Note that in both cases some compensating gate voltage is assumed to be applied, so that the ground spin states for the neutral (\mbox{$n=0$}) and charged (\mbox{$n=1$}) molecule are degenerate.
	}
	\label{fig3}
\end{figure}

To illustrate this point, let us consider the simplest model of a molecule for which such a situation arises: a molecule with \mbox{$S_0=3/2$} and \mbox{$S_1=2$} that exhibits only uniaxial magnetic anisotropy with~\mbox{$D\equiv D_0=D_1<0$}, and as previously, the contribution of only one vibrational mode is taken into account. The key feature of the energy spectrum of such a model molecule is that for both charge states the ground spin state(s), in each vibrational state, is formed by the state(s) characterized by the smallest projection of the spin along the $z$-axis, namely, \mbox{$\ket{0}$} and \mbox{$\ket{\pm1/2}$}.  
The corresponding energy spectrum in the absence of spin-vibron coupling~(\mbox{$\Lu\equiv\Lu_0=\Lu_1=0$}) is schematically depicted in Fig.~\ref{fig3}(a).

The situation changes as soon as~\mbox{$\Lu\neq0$}. In the limit where the charge-vibron coupling dominates (\mbox{$\Lu_n/\lambda\ll1$}) and for~\mbox{$\Lt_n=0$}, from Eqs.~(\ref{eq:Dnt_def})-(\ref{eq:Dnf_def}) one expects a positive correction~\mbox{$\approx2\lambda\Lu\hbar\omega$} to the otherwise negative uniaxial magnetic anisotropy constant~$D$ only in the charged state.
Then, for \mbox{$\Lu>|D|/(2\lambda\hbar\omega)$} one finds a reversal of the magnetic spectrum in the charged state, meaning that the states with the largest projection of the spin along the $z$-axis (\mbox{$\ket{\pm S_1}$}) again become lowest in energy, as one can see in Fig.~\ref{fig3}(b). Note at the same time that the magnetic spectrum in the neutral state remains approximately unaffected by coupling to molecular vibrations.
Importantly, the flip of the magnetic spectrum in only one charge state [as shown in Fig.~\ref{fig3}(b)] has a profound consequence for transport measurements, as transitions between the ground spin states of different charge states are no longer permitted by spin selection rules, see
Eq.~(\ref{eq:tunT_def}).\footnote{%
This point also justifies our deliberate choice of a model molecule which does not possess the transverse component of magnetic anisotropy. Did the molecule exhibit the transverse magnetic anisotropy, the ground spin state would consist of a superposition of pure $S_z$-projections, and thus, the transitions in question would be still allowed, though with lower weights.} 
This aspect will be further addressed in Sec.~\ref{sec:transport_blockade}.

On the other hand, at \mbox{$\Lu=|D|/(2\lambda\hbar\omega)$} all the spin states in the charged state become degenerate, so that the molecule effectively behaves  as if it was spin-isotropic. Actually, the spin-isotropic behavior should be observed already when \mbox{$\kB T,\Gamma\gtrsim (2S_1-1)|\Deff|$} with \mbox{$\Deff\approx D+2\lambda\Lu\hbar\omega$}.

%

\section{Transport characteristics}
\label{sec:transport}

As discussed in the previous section, the spin-vibron coupling can significantly influence the magnetic anisotropy of a molecule. 
In this section, we demonstrate how these effects manifest in the tunneling current through such a molecule in a transport setup as depicted in Fig.~\ref{fig1}(a). We focus on the two example molecules, for which we discussed the modified spectral properties in the previous section.

\subsection{Kinetic equations}
\label{sec:transport_theory}

In order to calculate the charge current through the molecule in the junction, we use a master equation approach derived from a real-time diagrammatic technique~\cite{Scholler1994,Koenig1996}. We start from the density matrix of the whole system and trace out the reservoir degrees of freedom. We are then left with the dynamics of the reduced density matrix with the elements~\mbox{$
	\rhoP_{\xi'}^{\xi}
	\equiv 
	\bra{\xi} \rhored \ket{\xi'}
$}. Here the states~\mbox{$\ket{\xi}\in\{\ket{\Chi_n}\otimes\ketv{\vartheta}\}$} denote the eigenstates of the vibrating molecule, when decoupled from the electronic reservoirs.
We are interested in transport in the stationary state and in a situation where the molecule is weakly coupled to the electrodes, \mbox{$\Gamma\ll \kB T$}. For this reason, we restrict our calculations to the sequential tunneling limit, where only first-order contributions in $\Gamma/(\kB T)$ are taken into account in the tunneling dynamics. Then, for the exemplary molecules discussed in Sec.~\ref{sec:spectrum}, the  dynamics of the diagonal elements of the reduced density matrix, \mbox{$\rhoP_\xi^\xi\equiv\rhoP_\xi$}, is governed by the Master equation
\begin{equation}\label{eq:MasterEq}
	\dfrac{\intd\rhoP_{\xi}}{\intd t} 
	=
	0
	= 
	\sum_{\xi\neq\xi^\prime} 
	\left(W_{\xi\xi^\prime}
	\rhoP_{\xi^\prime}-W_{\xi^\prime\xi}
	\rhoP_{\xi}\right)
	.
\end{equation}
The kernel $W_{\xi\xi^\prime}=\sum_{r=\text{S,D}}W_{\xi\xi^\prime}^{r}$ takes into account transition rates between molecular states due to (vibron-dependent) electron tunneling between the molecule and the source (\mbox{$r=\text{S}$}) or the drain (\mbox{$r=\text{D}$}). The elements of this kernel can be found employing Fermi golden rule.
For instance, the transition from a neutral state~\mbox{$\ket{\xi_0}=\ket{\Chi_0}\otimes\ketv{\vartheta}$} to a charged one~\mbox{$\ket{\xi_1}=\ket{\Chi_1}\otimes\ketv{\vartheta'}$} induced by tunneling of a single electron with spin~$\sigma$ from the $r$th electrode to the molecule occurs with the rate
\begin{equation}\label{eq:W_def}
	W_{\xi_1\xi_0}^{r\sigma}
	=
	\frac{\Gamma_\sigma^r}{\hbar}
	\big|\tunT_{\Chi_1\Chi_0}\big|^2
	\,
	\big|\FCcoef_{\vartheta'\vartheta}\big|^2
	f_r(\en_{\xi_1}-\en_{\xi_0})
	,
\end{equation}
with the coefficients $\tunT_{\Chi_1\Chi_0}$ and $\FCcoef_{\vartheta_1\vartheta_0}$ given by Eq.~(\ref{eq:tunT_def}) and Eq.~(\ref{eq:FCcoef_def}), respectively. 
It is important to emphasize that, while diagonal and off-diagonal elements of the reduced density matrix are decoupled in the example cases studied here, this is by no means a generally valid statement. 
In Appendix~\ref{app:coherences}, we show in detail how this decoupling occurs here, starting  from a full generalized kinetic equation that involves both the diagonal (\emph{occupation probabilities}) and the off-diagonal (\emph{coherences}) elements of the reduced density matrix of the molecule~$\rhored$~\cite{Braun2004,Weymann2005,Sothmann2010}.

We write the tunneling current through the device as the average of the currents through the tunnel barriers connecting the molecule to the drain~($I_\text{D}$) and the source~($I_\text{S}$), 
\begin{equation}\label{eq:current}
	I
	\equiv
	\frac{I_\text{D}-I_\text{S}}{2} 
	= 
	\frac{e}{2} \sum_{\xi,\xi'}\left(n_\xi-n_{\xi'}\right)\left(W_{\xi\xi^\prime}^{\text{D}}-W_{\xi\xi^\prime}^{\text{S}}\right)\rhoP_{\xi^\prime}
		,
\end{equation}
with the occupation probabilities~$\rhoP_{\xi^\prime}$ obtained from Eq.~\eqref{eq:MasterEq}. The variables $n_\xi$ take the value 0 or 1, depending on whether the molecule in state $\xi$ is neutral or charged, respectively. 

In what follows, we first give a general overview of features arising in transport spectroscopy due to the interplay of magnetic anisotropy and vibrations. Next, we present a specific case where transport characteristics of the device change radically if spin-vibron coupling is induced in the system.
In our discussion, we employ the two examples introduced in detail in Sec.~\ref{sec:spectrum}.

\subsection{Effect of the interplay of magnetic anisotropy and vibrations on transport characteristics}
\label{sec:spectroscopy}

\begin{figure*}[t!]
	\includegraphics[width=1\textwidth]{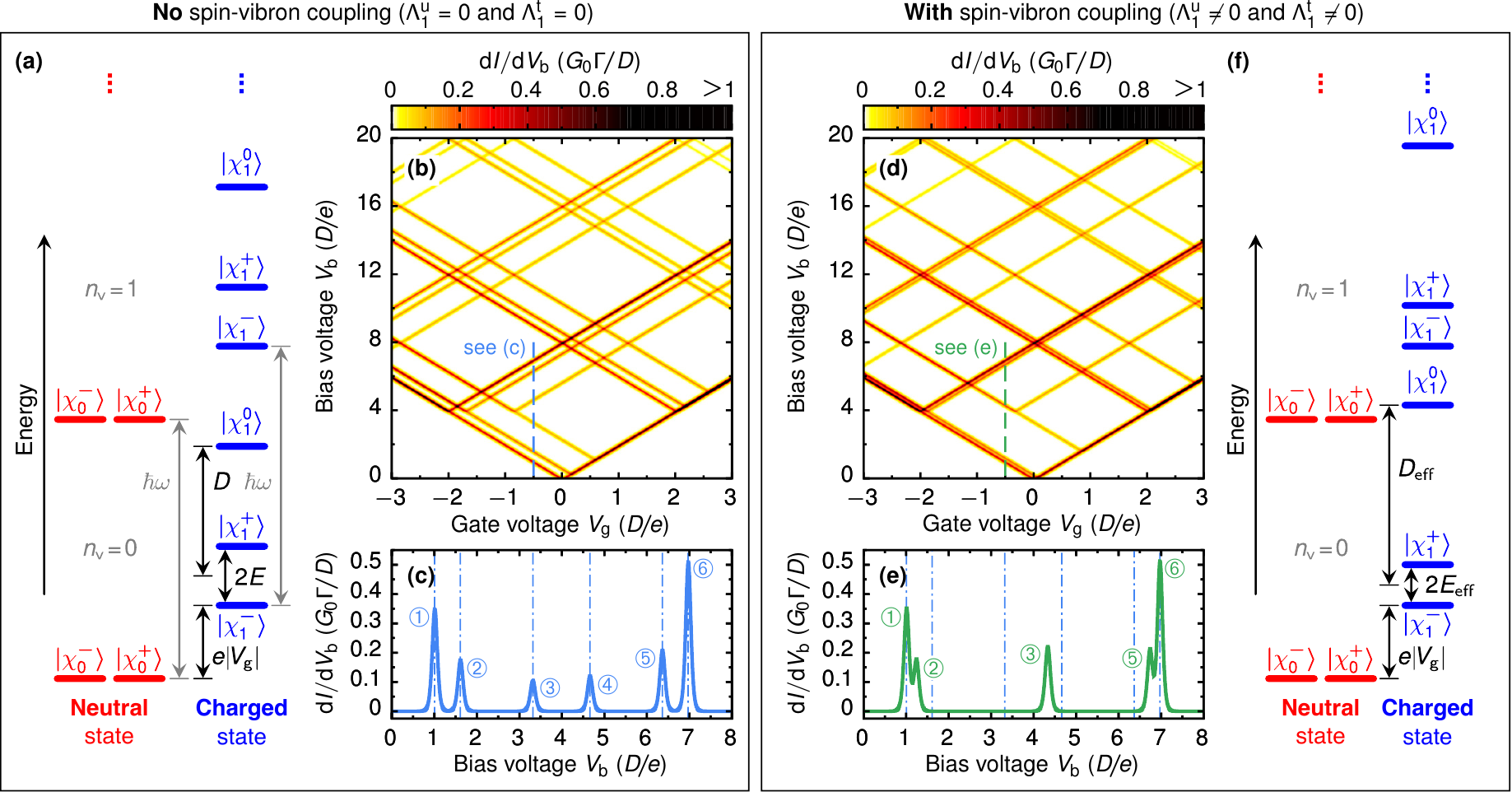}
	\caption{
	Effect of the charge- and spin-vibration couplings on transport characteristics of a tunnel junction containing a single molecule. 
	\emph{Left} (\emph{right}) box represents the case without (with) the spin-vibron coupling being included.
	(a,f) Schematic depiction of effective energy spectra for a molecule studied in Sec.~\ref{sec:spectroscopy}, where  two consecutive vibronic states~$\ketv{\nv}$ (for \mbox{$\nv=0,1$}) are shown.
	(b,d) Differential conductance~$\intd I/\intd \Vb$ as a function of gate~$\Vg$ and bias~$\Vb$ voltages for \mbox{$\lambda=1.5$} and nonmagnetic electrodes (\mbox{$P=0$}): (b) \mbox{$\Lu_1=\Lt_1=0$}, and (d) \mbox{$\Lu_1=0.05$} with \mbox{$\zeta=0.15$}.
	Here, \mbox{$G_0\equiv2e^2/h$} stands for the conductance quantum.
	(c) and~(e) Cross-sections of the density plots in (b) and~(d), respectively, taken at \mbox{$e\Vg/D=-0.5$} [that is, along the finely dashed lines in (b,d)], with the corresponding spectra given in~(a) and~(f).\textsuperscript{\ref{fn:Vg_comment}}
	Vertical thin dotted-dashed lines in (c,e), indicating the position of resonances in~(c), serve merely as a guide for the eye.
	Parameters assumed in calculations: \mbox{$\Gamma/D=0.01$}, \mbox{$\kB T/D=0.02$},  \mbox{$E/D=0.15$} and~\mbox{$\hbar\omega/D=4$} with \mbox{$D=500$}~$\mu$eV.
	}
	\label{fig4}
\end{figure*}

We will now investigate the impact of the spectral features for the model molecule discussed in Sec.~\ref{sec:E_D_spectrum} on the tunneling current through it. We therefore come back to the simple molecule with spin values~\mbox{$S_0=1/2$} and~\mbox{$S_1=1$}, whose spin-eigenstates in the neutral state are given by \mbox{$\ket{\chi_0^\pm}\equiv\ket{\pm1/2}$}, while in the charged state by
\mbox{$	
	\ket{\chi_1^0}
	=
	\ket{0}
$}
and
\mbox{$
	\ket{\chi_1^\pm}
	=
	\big(
	\ket{1}
	\pm
	\ket{-1}
	\big)
	/
	\sqrt{2}
	.
$}
Its effective energy spectrum (now including vibrational states) is schematically shown in Fig.~\ref{fig4}(a). 

Moreover, the following numerical results are obtained for realistic values of relevant parameters, that is, within the range of experimentally observed values, see \eg, Ref.~\cite{Burzur2014}.
Specifically, we assume that the coefficients characterizing intrinsic (static) magnetic anisotropy are~\mbox{$D=500$}~$\mu$eV and \mbox{$E/D=0.15$}, whereas the energy of a molecular vibrational mode is~\mbox{$\hbar\omega/D=4$}. 
We also note that except Sec.~\ref{sec:magnetic_electrodes}, we consider here nonmagnetic electrodes (\mbox{$P=0$}).

\subsubsection{No spin-vibron coupling}

To begin with, let us first consider the case where the molecule exhibits only the \textit{intrinsic} component of magnetic anisotropy, meaning that only charge-vibron (\mbox{$\lambda\neq0$}) but no spin-vibron coupling (\mbox{$\Lu_1=\Lt_1=0$}) is present. The corresponding spectrum together with the resulting differential conductance~\mbox{$\intd I/\intd\Vb$} is shown in the left box of Fig.~\ref{fig4}.%
\footnote{%
For the sake of simplicity and in order to enable easy comparison between the case without and with the spin-vibron coupling being present, we assume that some compensating gate voltage~$\Vg^\prime$ is always applied.  As a result,  at~\mbox{$\Vg=0$} the neutral doublet is degenerate with the charged ground state, see Fig.~\ref{fig4}(a,f).
\label{fn:Vg_comment}
} 
One can generally see that the spectroscopic features at low bias-voltage (\mbox{$e\Vb<2\hbar\omega$})%
\footnote{%
The factor `2' stems from the fact that the bias voltage~$\Vb$ is applied symmetrically to the electrodes, that is, \mbox{$\mu_{\text{S}(\text{D})}=\mu_0\pm e\Vb/2$}.
}
become duplicated whenever the bias voltage~$e\Vb$ exceeds twice the energy $\nv\hbar\omega$ (for~\mbox{$\nv=1,2,3\ldots$}) of the excited molecular vibrational state~$\ketv{\nv}$. 
The analysis of the position of resonances allows for extraction of the magnetic-anisotropy parameters~$D$ and~$E$, Eq.~(\ref{eq:H_spin}). For this purpose, in Fig.~\ref{fig4}(c) we plot  a representative cross-section from Fig.~\ref{fig4}(b) [see Fig.~\ref{fig4}(a) for the corresponding energy spectrum], and discuss the origin of resonances labeled~\mbox{\circon-\circsi}. These resonances essentially emerge due to transitions between different spin states, which follow the selection rules imposed thy the Clebsch-Gordon coefficients in Eq.~(\ref{eq:tunT_def}).
Specifically, the resonance~\circon is related to the ground-to-ground-state transitions \mbox{$\ket{\chi_0^\pm}\rightarrow\ket{\chi_1^-}$} \mbox{---note} that it is accompanied by a resonance mirrored with respect to \mbox{$\Vg=0$} representing transition in the opposite direction, \mbox{$\ket{\chi_1^-}\rightarrow\ket{\chi_0^\pm}$}. 
On the other hand, resonances~\circtw and~\circth correspond to the ground-to-excited-state transitions \mbox{$\ket{\chi_0^\pm}\rightarrow\ket{\chi_1^+}$} and \mbox{$\ket{\chi_0^\pm}\rightarrow\ket{\chi_1^0}$}, respectively. Consequently, from the relative position of resonances~\circon,~\circtw and~\circth one can deduce~$D$ and~$E$, as can be seen in Fig.~\ref{fig4}(a).

Furthermore, resonances~\circtw and~\circth can be observed only when a molecule becomes reduced (that is, it accepts one extra electron). Since the neutral state involves only a doublet state, no analogous resonances appear for the reverse process (oxidation).
All the resonances discussed so far stem from transitions between molecular spin states belonging to the ground molecular vibrational state, that is, for~\mbox{$\nv=0$}. However, when also transitions between different vibrational states are energetically permitted, the excited-to-excited-state transitions become visible for the oxidation process. Resonances representing such transitions are, for instance, those labeled as~\circfo~(for \mbox{$\ket{\chi_1^0}\otimes\ketv{0}\rightarrow\ket{\chi_0^\pm}\otimes\ketv{1}$}) and~\circfi~(for \mbox{$\ket{\chi_1^+}\otimes\ketv{0}\rightarrow\ket{\chi_0^\pm}\otimes\ketv{1}$}). The characteristic property of these resonances, which can be seen in Fig.~\ref{fig4}(b), is that they do not continue to resonance~\circon. Instead, they terminate at resonances associated with single-electron-tunneling-in transitions that lead to occupation of relevant excited states, namely, resonances~\circfo and~\circfi  terminate at~\circth and~\circtw, respectively.
Finally, the last pronounced resonance~\circsi in Fig.~\ref{fig4}(c) arises owing to transitions between ground spin states of two neighboring vibrational states, that is, \mbox{$\ket{\chi_1^-}\otimes\ketv{\nv}\rightarrow\ket{\chi_0^\pm}\otimes\ketv{\nv^\prime}$} with \mbox{$\nv^\prime-\nv=1$}. Since the dominating contribution comes from the transition between the ground~(\mbox{$\nv=0$}) and first excited (\mbox{$\nv^\prime=1$}) vibrational states, resonance~\circsi in Fig.~\ref{fig4}(b) reaches resonance~\circon.  Note that from the position of~\circsi one can easily determine the energy of the vibrational mode, see Fig.~\ref{fig4}(a).

The physical origin of resonances visible at larger bias voltage (\mbox{$e\Vb\geqslant2\hbar\omega$}) can be understood using the same arguments as above. The only difference is now that transitions take place between states with different numbers of molecular vibrational excitations.
Moreover, the intensity of equivalent resonances (that is, associated with the same type of spin transitions but occurring between different vibrational states) is attenuated. 
This effect is governed by the Franck-Condon factors, Eq.~(\ref{eq:FCcoef_def}), which basically put a weight on transition rates determined by the nuclear wave function overlap between the various vibrational states of the molecules~\cite{Koch2006Nov,Seldenthuis2008}.

\subsubsection{Spectroscopic signatures of spin-vibron coupling}
\label{sec:spectroscopy_spin-vib}

The situation changes if also the spin-vibron coupling becomes active, which is illustrated in the right box of Fig.~\ref{fig4}, with the density plot of the differential conductance~\mbox{$\intd I/\intd\Vb$} given in panel~(d) and a relevant cross-section for \mbox{$e\Vg/D=-0.5$} shown in panel~(e).

The position of resonances~\circtw and~\circth associated with the value of the uniaxial and transverse component of magnetic anisotropy, respectively, is shifted; compare in Fig.~\ref{fig4} panel~(c) for~\mbox{$\Lu_1=0$} with panel~(e) for~\mbox{$\Lu_1\neq0$}. In particular, resonance~\circth moves towards larger bias voltages (\mbox{$\Deff>D$}), while for resonance~\circtw the opposite behavior is observed (\mbox{$\Eeff<E$}), see the pertinent energy spectrum schematically shown in Fig.~\ref{fig4}(f).
Physically, it corresponds to increasing the energy barrier for spin reversal (determined by~$\Deff$), while reducing the effect of under-barrier transitions (introduced by~$\Eeff$).
Moreover, we also note that resonance~\circfo from Fig.~\ref{fig4}(c) is absent in Fig.~\ref{fig4}(e). The underlying transition does not arise in the present situation, because the energy of the state~\mbox{$\ket{\chi_0^\pm}\otimes\ketv{1}$} is smaller than that for \mbox{$\ket{\chi_1^0}\otimes\ketv{0}$}, compare panels~(a) and~(f) in Fig.~\ref{fig4}.
In experiment, measuring the shifts~\mbox{$\Delta D=\Deff-D$} and~\mbox{$\Delta E=\Eeff-E$} would allow for estimating the spin-vibron coupling strengths~$\Lu_1$ and~$\Lt_1$ by means of Eqs.~(\ref{eq:DeltaD_def})-(\ref{eq:DeltaE_def}).

Moreover, if one could control and increase further the strength of the spin-vibron coupling, it should in principle be possible to diminish the gap between states~$\ket{\chi_1^\pm}$ beyond the detection limit set here predominantly by temperature~$T$. 
One of promising ways to achieve this goal may be to tune the coupling \via stretching of the molecule embedded in a mechanically controllable break junction. Realistic changes of the coupling strength obtained with this method are expected to be  of the order of a few percent~\cite{Adamczewska_cooment}. It is also for this reason that we chose to show the example in the right box of Fig.~\ref{fig4} and to not consider the case where~$E_\text{eff}$ can get fully suppressed (up to $\Gamma$ and below) \via the spin-vibron coupling.   
Nevertheless, for some specific molecules it may still be possible to completely switch off the transverse component of magnetic anisotropy in this way. 

\subsubsection{Asymmetry effect of spin-vibron coupling}
\label{sec:Effect_of_zeta}

\begin{figure}[t]
	\includegraphics[scale=1]{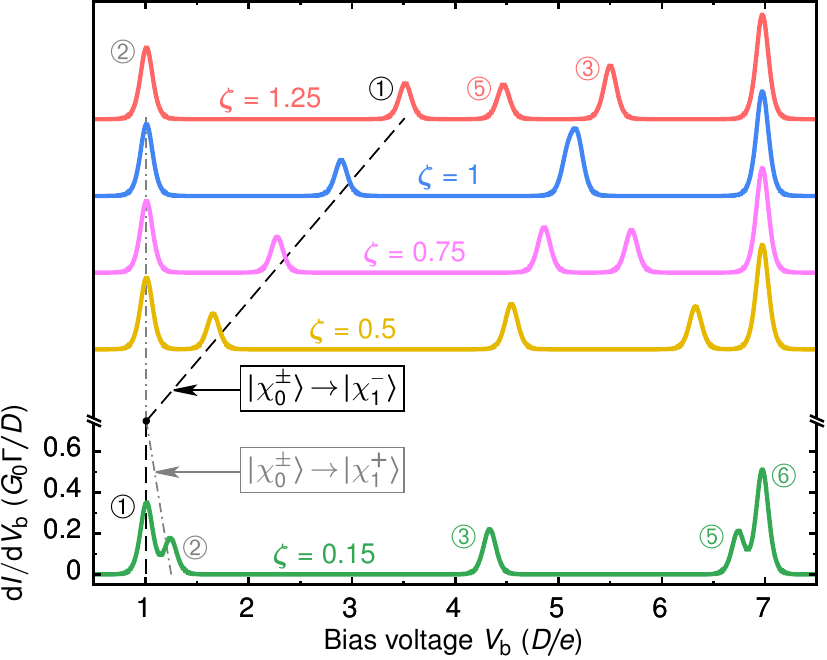}
	\caption{
	Influence of the asymmetry between the transverse and the uniaxial component of the spin-vibron coupling (quantified by~\mbox{$\zeta=\Lt_1/\Lu_1$}) on the differential conductance shown for indicated values of~$\zeta$.
	For clarity, curves for~\mbox{$\zeta>0.15$} are shifted vertically, with the bottom curve for~\mbox{$\zeta=0.15$} being identical to that presented in Fig.~\ref{fig4}(e).
	Note that, as previously, some compensating gate voltage is applied to fix the position of the (left-most) resonance corresponding to the ground-to-ground-state transitions, and thus, to enable easy comparison of different curves.
	Other parameters are taken the same as in the right box of Fig.~\ref{fig4}.
	}
	\label{fig5}
\end{figure}

In the previous subsection, we  made the assumption that the ratio of the transverse to the uniaxial component of the spin-vibron coupling, \mbox{$\zeta=\Lt_1/\Lu_1$}, is approximately equal to \emph{$\zeta\approx E/D=0.15$}. However, in real systems this condition does not necessarily have to be satisfied. For this reason, here we discuss how the asymmetry between different components of spin-vibron coupling (quantified by~$\zeta$) becomes visible in transport spectroscopy.

First of all, recall from Sec.~\ref{sec:E_D_spectrum} that while the correction~$\Delta D$ to the uniaxial magnetic anisotropy [see Eq.~(\ref{eq:dD_magnitude})] only weakly depends on~$\zeta$, in the case of the correction~$\Delta E$ to the transverse magnetic anisotropy [see Eq.~(\ref{eq:dE_magnitude})] this dependence is linear. As a result, the value of~$\zeta$ should more significantly affect transport features associated with the energy scale~$2\Eeff$ rather than the ones associated with~$\Deff$. In particular, the position of resonances \mbox{\circon{\,}--\,\circth} and \circfi in Fig.~\ref{fig4} discussed in the former subsection are thereby 
modified.\footnote{%
Experimentally, it might be difficult to discern the swap between resonances \circon and \circtw, discussed in the following, and it might therefore seem as if only resonance \circtw got affected.}

In Fig.~\ref{fig5} we analyze how the differential conductance plotted in Fig.~\ref{fig4}(e) [shown here for reference as the green curve for~\mbox{$\zeta=0.15$}] depends on the value of~$\zeta$ ---note that the coupling parameter~$\Lu_1$ is fixed in  the present considerations (\mbox{$\Lu_1=0.05$}).
As discussed in Sec.~\ref{sec:E_D_spectrum}, the relation between~$\Lu_1$ and~$\Lucrit$ [see Eq.~(\ref{eq:Lu_crit})] determines the ground spin state of a charged molecule, namely: $\ket{\chi_1^-}$ if \mbox{$\Lu_1<\Lucrit$}, and $\ket{\chi_1^+}$ if \mbox{$\Lu_1>\Lucrit$}, which has been graphically depicted in Fig.~\ref{fig2}.
Importantly, when increasing~$\zeta$ the critical value~$\Lucrit$  is effectively diminished. 
Therefore, one finds that at fixed $\Lu_1$, $\ket{\chi_1^-}$ is the ground state for \mbox{$\zeta\lesssim\zeta^\ast$}, while $\ket{\chi_1^+}$ is the ground state for \mbox{$\zeta\gtrsim\zeta^\ast$}, with
\begin{equation}
	\zeta^\ast
	=
	\frac{
	E
	}{
	2\lambda\Lu_1
	\big(1+\Lu_1/\lambda\big)
	\hbar\omega
	}
	.
\end{equation}
For the parameters used in Fig.~\ref{fig5}, one finds \mbox{$\zeta^\ast\approx0.24$}.
In consequence, one expects that: (i) \mbox{$0<\Eeff<E$} for \mbox{$\zeta\lesssim\zeta^\ast$}, and in particular, \mbox{$\Eeff\approx E$} for negligibly small~$\zeta$; (ii) \mbox{$\Eeff<0$} for \mbox{$\zeta\gtrsim\zeta^\ast$}, and additionally if  \mbox{$\zeta>2\zeta^\ast$} one finds \mbox{$|\Eeff|>E$}.
These distinctive regimes translate into specific shifts of characteristic resonances in the differential conductance, see Fig.~\ref{fig5}. 
To illustrate this point, as an example, we have schematically indicated there with thin lines the evolution of resonances marked as \circon (dashed line) and \circtw (dotted-dashed line), corresponding to transitions \mbox{$\ket{\chi_0^\pm}\rightarrow\ket{\chi_1^-}$} and \mbox{$\ket{\chi_0^\pm}\rightarrow\ket{\chi_1^+}$}, respectively.
For large~$\zeta$ (that is, for \mbox{$\zeta\gtrsim2\zeta^\ast$}) the two resonances are well separated, which would allow for a more accurate readout of excitation energies.

\subsubsection{Potential of magnetic electrodes}
\label{sec:magnetic_electrodes}

\begin{figure}[t]
	\includegraphics[scale=1]{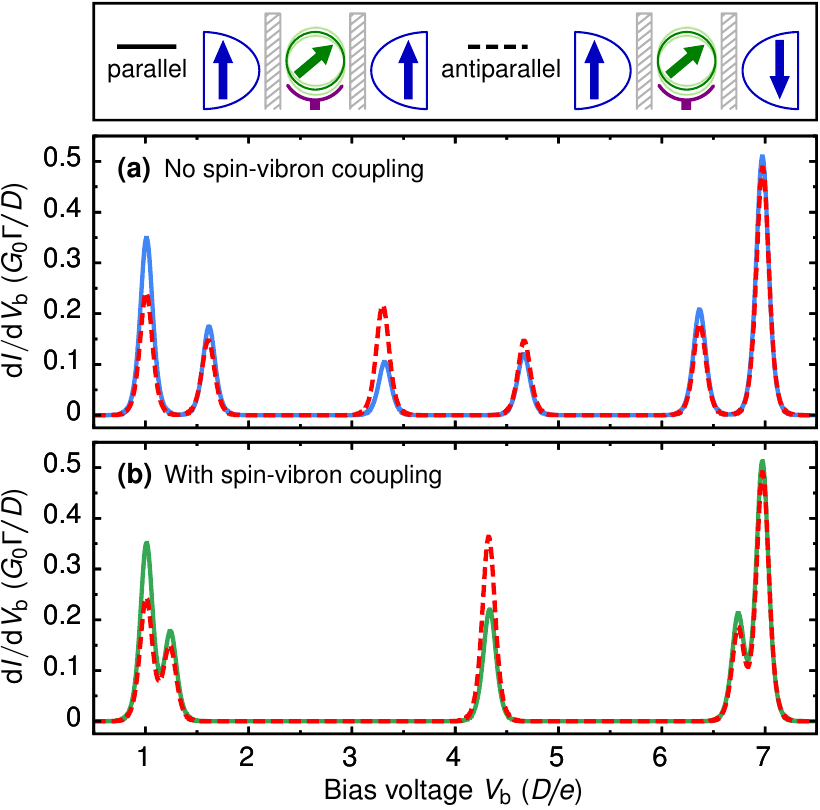}
	\caption{
	Selective effect of two different collinear magnetic configurations of the device [that is, for \emph{parallel} (solid lines) and  \emph{antiparallel} (dashed lines) relative orientation of the spin moments in the electrodes (for \mbox{$P=0.5$})] on differential conductance~\mbox{$\intd I/\intd\Vb$}.
	Note that solid lines in panels~(a) and~(b) are identical to those in panels~(c) and~(e) of~Fig.~\ref{fig4}, respectively, obtained for nonmagnetic electrodes (\ie, for \mbox{$P=0$}).
	All remaining parameters as in Fig.~\ref{fig4}.
	}
	\label{fig6}
\end{figure}

Finally, we note that the advantage of using a magnetic junction is that one can selectively enhance or decrease resonances.
So far, we have concentrated exclusively on transport characteristics of the device in the case of \emph{nonmagnetic} electrodes, see Fig.~\ref{fig4} and Fig.~\ref{fig5}. Noteworthily, when using \emph{magnetic} electrodes, by switching the device from the parallel into the antiparallel magnetic configuration, one can adjust the intensity of certain resonances.

In Fig.~\ref{fig6} we compare cross-sections of the differential conductance at a fixed gate voltage obtained by changing the relative orientation of spin moments of the source and the drain from parallel (solid lines) to antiparallel (dashed lines).  
Importantly, note that the solid lines for the parallel magnetic configuration are in fact identical to those calculated in Figs.~\ref{fig4}(c,e) for nonmagnetic electrodes.
It can be seen that while a majority of resonances is only weakly affected by the change of the magnetic configuration, two resonances visibly react to it: resonance~\circth becomes more pronounced and the intensity of resonance~\circon gets diminished. In the latter case, by reducing the disproportion between the heights of resonances~\circon and~\circtw, one expects to better resolve the merging of the two resonances when for example $\zeta$ or $\Lu_1$ are changed as discussed in the previous section.
The mechanism underlying this effect stems from the spin-asymmetry of the tunnel coupling of a molecule to the drain and source electrodes, given in the end of Sec.~\ref{sec:theory}. It basically leads to unequal occupation probabilities of the neutral-doublet states~$\ket{\chi_0^-}$ and~$\ket{\chi_0^+}$, which affect, in turn, the current flowing through the molecule, Eq.~(\ref{eq:current}).

\subsection{Vibrationally induced spin blockade in transport}
\label{sec:transport_blockade}

\begin{figure}[t]
	\includegraphics[scale=1]{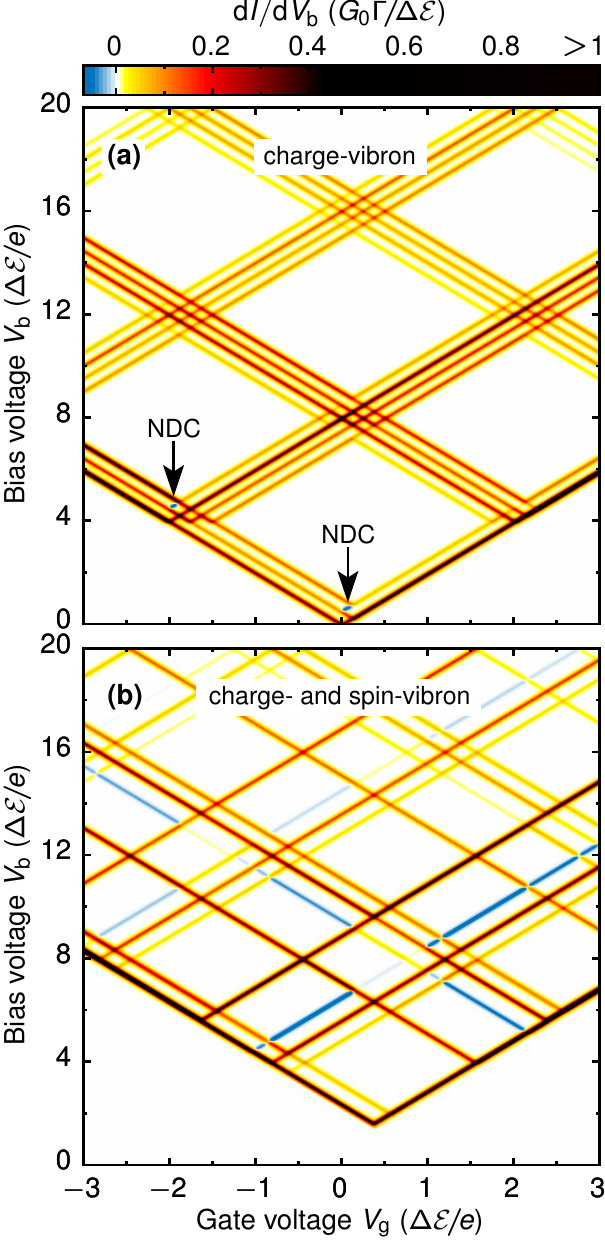}
	\caption{
	The effect of spin blockade in transport induced by the reversal of the magnetic spectrum due to the spin-vibron coupling.
	Differential conductance~$\intd I/\intd\Vb$ of a device based on a model molecule with~\mbox{$S_0=3/2$} and \mbox{$S_1=2$}, whose energy spectra are schematically shown in Fig.~\ref{fig3}, is plotted as a function of the gate~$\Vg$ and bias~$\Vb$ voltage for: 
	(a) \mbox{$\Lu_0=\Lu_1=0$}, and (b) \mbox{$\Lu_0=\Lu_1=0.05$}. 
	Note that the energy unit \mbox{$\Delta\en=4|D|$} corresponds to the difference between energies of the spin states \mbox{$\ket{0}$} and \mbox{$\ket{\pm 2}$}  of the charged molecule without spin-vibron coupling, see also the right side of Fig.~\ref{fig3}(a).
	NDC stands here for `negative differential conductance'.
	The  other parameters are \mbox{$\Gamma/\Delta\en=0.01$}, \mbox{$P=0$}, \mbox{$\lambda=1.5$}, \mbox{$\kB T/\Delta\en=0.02$}, \mbox{$E=0$}, \mbox{$\hbar\omega/\Delta\en=4$} with \mbox{$\Delta\en=4|D|=500$}~$\mu$eV.
	}
	\label{fig7}
\end{figure}

Finally, we show that the reversal of the magnetic spectrum in a large-spin molecule due to the coupling of spin and charge to molecular vibrations,  non-trivially manifests in transport spectroscopy.  As already announced in Sec.~\ref{sec:barrier_flip}, it can lead to the occurrence of a spin-blockade in transport, which we investigate in the present section.
For this purpose, we employ the minimal model of a molecule capable of supporting such an effect, characterized by spins \mbox{$S_0=3/2$} and \mbox{$S_1=2$}, which exhibits only a (negative) uniaxial component of magnetic anisotropy, here assumed to be~\mbox{$D\equiv D_0=D_1=-125$}~$\mu$eV. The relevant magnetic spectrum of such a molecule is schematically shown in Fig.~\ref{fig3}.
For conceptual simplicity, we again include only one vibrational mode with energy~\mbox{$\hbar\omega=2$ meV}, and take the coupling parameters \mbox{$\lambda=1.5$} and \mbox{$\Lu\equiv\Lu_0=\Lu_1=0.05$}, while consistently neglecting the transverse component of the coupling, that is, \mbox{$\Lt_0=\Lt_1=0$}. For other parameters see the caption of Fig.~\ref{fig7}.

We show the differential conductance of this model system in Fig.~\ref{fig7} for both cases without [panel~(a)] and with [panel~(b)] the spin of the molecule being coupled to its vibrations. 
In the former situation [panel (a)], one can see that the behavior of the differential conductance as a function of bias and gate voltages qualitatively resembles that for the molecule analyzed in Fig.~\ref{fig4}(b), but with more transitions since the molecule is characterized by a larger spin. The observed resonances can be attributed to specific transitions between different charge states~\mbox{$\ket{M_0}\otimes\ketv{\nv}$} and~\mbox{$\ket{M_1}\otimes\ketv{\nv^\prime}$} [see Fig.~\ref{fig3}(a)] that satisfy the spin selection rule~\mbox{$|M_1-M_0|=1/2$}.
The only new features are some (blue) spots of negative differential conductance (NDC, marked by arrows), which signify a reduction of transport. The NDC arises when the molecule gets trapped in the excited doublet state for~\mbox{$n=0$} (\ie, the state $\ket{\pm 3/2}$), before the transition to the highest-in-energy doublet state for~\mbox{$n=1$} (\ie, the state $\ket{\pm 2}$) becomes energetically permitted by application of a bias voltage. This NDC is possible since the energy required for the transitions  \mbox{$\ket{\pm 1/2}\rightarrow\ket{\pm 1}$} and \mbox{$\ket{\pm 1}\rightarrow\ket{\pm 3/2}$} is the same, while the excitation energy for \mbox{$\ket{\pm 3/2}\rightarrow\ket{\pm 2}$} is two times larger. See also the spectra in Fig.~\ref{fig3}(a) for clarification.

Also in the presence of spin-vibron coupling [see Fig.~\ref{fig7}(b)], extended regions of NDC are visible. However, what is more striking is that at  low bias voltage, \mbox{$eV_\text{b}\lesssim2\Delta\en$}, transport is \textit{fully suppressed}. The reason for this is that for the  present, purposefully chosen set of parameters, one finds from Eqs.~(\ref{eq:Dnt_def})-(\ref{eq:Dnf_def}) that while the uniaxial magnetic anisotropy constant for the neutral state remains approximately the same, in the charged state the new effective anisotropy constant \mbox{$\Deff\approx D+2\lambda\Lu\hbar\omega$} is positive. As a result, an energy barrier for spin reversal in the charged state forms, as illustrated in Fig.~\ref{fig3}(b).
Most noticeably, the reversal of the magnetic spectrum entails that only transitions between ground and excited spin states (of the neutral and the charged molecule, respectively)  are allowed by spin selection rules.

\section{Summary and conclusions}
\label{sec:conclusions}

The main purpose of this paper was to investigate the effect of the coupling of molecular vibrations  to the charge and spin of a molecule on magnetic properties of such a molecule.
By deriving the effective giant-spin Hamiltonian, Eq.~(\ref{eq:H_spin_2}), we have found that these vibronic couplings result in modifications of the magnetic anisotropy parameters of the molecule, along both the uniaxial [see Eqs.~(\ref{eq:Dnt_def})-(\ref{eq:Dnf_def})] and transverse [see Eqs.~(\ref{eq:Ent_def})-(\ref{eq:Cnf_def})] directions, by inducing additional magnetic anisotropy components.
Depending on the intrinsic magnetic anisotropy of the molecule, its vibrational energy and the coupling strength to its spin, this interaction can lead to diverse effects ranging from enhancing to quenching or even inverting different components of the magnetic anisotropy.

In order to illustrate how the effect of spin-vibron coupling  manifests in transport spectroscopy, we have considered a device consisting of a single magnetic molecule inserted in a capacitively gated three-terminal junction.  We have perturbatively calculated stationary transport in first order of the tunnel-coupling  using a real-time diagrammatic technique.
In our calculations, we have paid particular attention to justify the conditions  under which coherent superpositions between molecular states (represented by the off-diagonal components of the reduced density matrix of a molecule) play no role for transport.

Our results show that the modulations of the magnetic anisotropy can lead to distinct effects in the differential conductance. In particular, in certain molecular regimes even  a blockade of transport can occur.
We expect that the effects under discussion, stemming from the spin-vibron coupling, should be observable especially in molecules based on individual metallic/magnetic ions, such as, Co-based complexes~\cite{Parks2010} or metal complexes derived from phthalocyanine (based on single ions of Cu, Mn, Fe, Ni)~\cite{Mugarza_Nat.Commun.2/2011,Urdampilleta_Nat.Mater.10/2011,Rakhmilevitch2014}. In such molecules, their magnetic core is particularly sensitive to changes of the crystal field of surrounding ligands associated with molecular vibrations. For instance, such a mechanism has been proposed~\cite{Ruiz2012} to explain the experiment by Parks~\etal~\cite{Parks2010}.

In general, junctions containing a single magnetic molecule owe their interest to envisioned applications of such systems as information storing and processing devices. In this context, the analysis conducted in this paper provides an insight on how to harness molecular vibrations to control the magnetic anisotropy. We show that it constitutes a possible mechanism to enhance a magnetic bistability of such molecules, which is a necessary requirement for a binary memory element. For instance, by mechanically stretching the junction or by deforming the molecule using other means, the energy of the vibrational modes, as well as, the coupling strength to the molecular spin can be tuned to increase the energy barrier for spin reversal while reducing the effect of magnetization tunneling under the barrier. Consequently, our results indicate a way to improve the robustness of spintronics devices based on single magnetic molecules.

\acknowledgments

We thank  Ma\l{}gorzata Ademczewska-Wawrzyniak for fruitful discussion.
Financial support  from the Knut and Alice Wallenberg Foundation (J.S. and M.M.)  and the Swedish VR (J.S.) is acknowledged. 
M. M. also acknowledges financial support from the Polish Ministry of Science and Higher Education through a Iuventus Plus project (IP2014 030973) in years 2015-2017 and a young scientist fellowship (0066/E-336/9/2014).

\appendix
\section{Effect of coherences on sequential-tunneling transport}
\label{app:coherences}

In Sec.~\ref{sec:transport_theory}, we have introduced the Master equation for the diagonal elements of the reduced density matrix of the molecule $\rhoP_\xi^\xi$, see Eq.~\eqref{eq:MasterEq}. However, only in special, yet relevant cases, the dynamics of the diagonal and off-diagonal (\ie, \mbox{$\rhoP_{\xi^\prime}^\xi$} for \mbox{$\xi\neq\xi^\prime$}) elements of the reduced density matrix decouple. In the following, we explain why the example cases studied in the present paper can indeed be treated with a simple Master equation as given in  Eq.~\eqref{eq:MasterEq}.

More generally, the kinetic equation for a weakly coupled molecule in the stationary regime reads
\begin{equation}\label{eq:MasterEqFull}
	\dfrac{\intd\rhoP_{\xi_2}^{\xi_1}}{\intd t} 
	=
	0
	= 
	-\frac{i}{\hbar}
	(\en_{\xi_1} - \en_{\xi_2})
	\rhoP_{\xi_2}^{\xi_1} 
	+
	\sum_{\xi_1^\prime\xi_2^\prime} 
	W_{\xi_2^{}\xi_2^\prime}^{\xi_1^{}\xi_1^\prime}
	\rhoP_{\xi_2^\prime}^{\xi_1^\prime}
	.
\end{equation}%
The first term in the right hand side of the equation above represents the intra-molecule dynamics, and it only plays a role for coherences~(\mbox{$\xi_1\neq\xi_2$}), vanishing for occupation probabilities~(\mbox{$\xi_1=\xi_2$}). The second term, on the other hand, involves transitions between molecular states induced by tunneling of electrons between the molecule and electrodes. 
These processes are captured \via the generalized transition rates~$W_{\xi_2^{}\xi_2^\prime}^{\xi_1^{}\xi_1^\prime}$, which can be evaluated using a real-time diagrammtic technique~\cite{Koenig1996}. For explicit rules for the diagrammatic evaluation of these rates, see, \eg, Appendix~B in Ref.~\cite{Koenig1996} or Appendix~A in Ref.~\cite{Sothmann2010}.

\subsubsection{Energy splitting}

In principle, coherent superpositions between two molecular states~$\ket{\xi}$ and~$\ket{\xi^\prime}$, represented by the off-diagonal terms~$\rhoP^{\xi}_{\xi^\prime}$ of the reduced density matrix~$\rhored$, play a role in the sequential-tunneling regime only if
\mbox{$
	|\en_{\xi}-\en_{\xi^\prime}|
	\lesssim
	\Gamma
$}~\cite{Sothmann2010}.
The reason for this is the following:
when the energy separation~\mbox{$|\en_{\xi}-\en_{\xi^\prime}|$}  significantly exceeds the tunnel-broadening~$\Gamma$ of these states, there is a zeroth order term in the perturbation expansion in~$\Gamma/(\kB T)$ to the Master equation. As a result, the first term on the right hand side of Eq.~\eqref{eq:MasterEqFull} ---being the only contribution in this order--- has to equal zero.
In this regime, coherences~$\rhoP^{\xi}_{\xi^\prime}$ can thus be neglected and Eq.~(\ref{eq:MasterEq}) is a valid approximation describing the dynamics of the molecule's occupation probability.
In general, how to treat coherences in systems where  states with~\mbox{$|\en_{\xi}-\en_{\xi^\prime}|\approx\Gamma$} occur, depends on the specific properties of the studied molecule. 
For the case of molecules with uniaxial and transverse anisotropy, studied in this paper,  it means that only coherences between  states coupled by transverse magnetic anisotropy [see the second term of Eq.~(\ref{eq:H_spin})], which are either degenerate or slightly split, can have an impact on the dynamics.

\subsubsection{Spin-selection rules}
 
Let us first concentrate on molecules with~spins~$S_n$ exhibiting only uniaxial  magnetic anisotropy~(\mbox{$E_n=0$}). In~such a case, the spin states of the molecule correspond to pure \mbox{$S_z$-spin} projections~$\ket{M_n}$ (with $M_n=-S_n,-S_n+1,\ldots,S_n-1,S_n$), see Sec.~\ref{sec:model_mol}. 
Following the previous subsection, we need to examine the behavior of \textit{degenerate} (time-reversed) states \mbox{$\ket{\!\pm |M_n|}$}. However, in the present situation, transitions between the diagonal~($\rhoP_\xi$) and off-diagonal~($\rhoP_{\xi^\prime}^\xi$) elements of the reduced density matrix~$\rhored$ are forbidden due to spin conservation. Consequently, the dynamics of~$\rhoP_\xi$ and~$\rhoP_{\xi^\prime}^\xi$ decouples and Eq.~\eqref{eq:MasterEq} is valid in the sequential-tunneling limit.

On the other hand, if also the~transverse  component of intrinsic magnetic anisotropy~(\mbox{$E_n\neq0$}) exists, the spin states~$\ket{\Chi_n}$ become composed of an admixture of states~$\ket{M_n}$ belonging to one of two otherwise uncoupled, time-reversed sets~\cite{Misiorny2015}.

For a \emph{half-integer} spin~$S_n$, the states~\mbox{$\ket{\Chi_n}=\ket{\!\pm |M_n|}$} form Kramers' doublets.  Their degeneracy cannot be lifted by the presence of a transverse anisotropy \mbox{---indeed,} the transverse anisotropy couples only those states for which $\Delta M_z$ is an integer  multiple  of 2. In practice, this means that only spin-transitions between states from different doublets are enabled when \mbox{$E_n\neq0$}. Such states, however, have a large energy splitting due to the uniaxial anisotropy, \mbox{$D\gg\Gamma$}, and coherences between them are therefore suppressed in first-order transport.

In contrast, for an \emph{integer} spin~$S_n$, previously degenerate states \mbox{$\ket{\Chi_n}=\ket{\pm M_n}$} become coupled by the transverse magnetic anisotropy. They thereby get split by an energy~$\Delta$, as shown for~\mbox{$S_n=1$} in the right side of Fig.~\ref{fig1}(b) where~\mbox{$\Delta=2E$} (with \mbox{$E\equiv E_n$}). 
For~\mbox{$S_n>1$}, this energy splitting~$\Delta$ can even be significantly smaller than~$E_n$. Consequently, for small (effective) transverse anisotropies with \mbox{$\Delta\lesssim \Gamma$}, the contribution of coherences to the molecule dynamics might be relevant. 
Nevertheless, whether coherences in the end really contribute or not, still depends on the specific transport setup. Below, we discuss in more detail the relevant example of a molecule with~\mbox{$S_0=1/2$} and~\mbox{$S_1=1$}, already introduced in Fig.~\ref{fig1}(b) and discussed in Secs.~\ref{sec:E_D_spectrum} and~\ref{sec:spectroscopy}.

\subsubsection{Example of a spin-1 molecule}

In this subsection, we demonstrate that coherent superpositions between the magnetic states~$\ket{\chi_1^+}$ and~$\ket{\chi_1^-}$ must be included, if~$2\Eeff\lesssim\Gamma$ and the electronic contacts are magnetic and differently polarized. 
Since, due to the large energy splitting $\hbar\omega$, no coherences between different vibrational states arise, we consider in the following the conceptually simplest case of a \emph{static} molecule with states \mbox{$\ket{\xi_n}\equiv\ket{\Chi_n}$}, where \mbox{$\ket{\Chi_0}\in\big\{\ket{\chi_0^\pm}\big\}$} and $\ket{\Chi_1}\in\big\{\ket{\chi_1^0},\ket{\chi_1^\pm}\big\}$, as defined in Sec.~\ref{sec:E_D_spectrum}.
Thus, the reduced density matrix~$\rhored$ of the molecule in matrix representation reads as
\begin{equation}\label{eq:rhored_ex}
\renewcommand{\arraystretch}{1.5}
	\rhored
	=
	\begin{pmatrix}
	\rhoP_{\chi_0^-} & 0 & 0 & 0 & 0
	\\
	0 & \rhoP_{\chi_0^+} & 0 & 0 & 0
	\\
	0 & 0 & \rhoP_{\chi_1^-} & 0 & \rhoP_{\chi_1^-}^{\chi_1^+}
	\\
	0 & 0 & 0 & \rhoP_{\chi_1^0} & 0
	\\
	0 & 0 & \rhoP_{\chi_1^+}^{\chi_1^-} & 0 & \rhoP_{\chi_1^+}
	\end{pmatrix}
	,
\end{equation}
with the diagonal elements representing the probabilities, and two off-diagonal elements
capturing the coherent superpositions between states~$\ket{\chi_1^+}$ and~$\ket{\chi_1^-}$.

Using diagrammatic rules for the evaluation of the kernel, see, \eg, Refs.~\cite{Koenig1996,Sothmann2010} and Appendix~\ref{app:exp} for explicit expressions, we can write down the full set of Master equations~(\ref{eq:MasterEqFull}) for the entries of the reduced density matrix, Eq.~(\ref{eq:rhored_ex}), in first order in the tunnel-coupling.
An intuitive physical understanding can be gained by expressing them in the form of Bloch equations.
For this purpose, we introduce a pseudospin vector~$\vSigmaog$ for the two lowest-in-energy (ground) spin states~$\ket{\chi_1^+}$ and~$\ket{\chi_1^-}$ of the charged molecule (see the left side of Fig.~\ref{fig2}). It is defined as
\begin{equation}\label{eq:Sigma1_def}
\renewcommand{\arraystretch}{1.75}
	\vSigmaog
	= 
	\begin{pmatrix}
	\Sigmaog^{x}
	\\
	\Sigmaog^{y}
	\\
	\Sigmaog^{z}
	\end{pmatrix}
	= 
	\dfrac{1}{2}
	\begin{pmatrix}
	\rhoP_{\chi_1^+}^{\chi_1^-} + \rhoP_{\chi_1^-}^{\chi_1^+}
	\\
	i \Big[ \rhoP_{\chi_1^+}^{\chi_1^-} - \rhoP_{\chi_1^-}^{\chi_1^+}\Big]
	\\
	\rhoP_{\chi_1^+} -  \rhoP_{\chi_1^-}
	\end{pmatrix}
	,
\end{equation}
and analogously for the degenerate neutral state
%
\begin{equation}\label{eq:Sigma0_def}
	\Sigma_0^z
	=
	\big(\rhoP_{\chi_0^+} - \rhoP_{\chi_0^-}\big)/2\ .
\end{equation}
%
In addition, the probabilities to find the molecule in the neutral state, $\rhoP_0$,  in the charged ground state, $\rhoPog$, or in the excited charged state, $\rhoP_\oe$, are given by
\begin{equation}\label{eq:prob_def}
\left(\begin{array}{c}
	\rhoP_0 \\
	\rhoPog\\
	\rhoP_\oe
\end{array}\right)	
	=
\left(\begin{array}{c}	
	\rhoP_{\chi_0^-} + \rhoP_{\chi_0^+}\\
	\rhoP_{\chi_1^-} + \rhoP_{\chi_1^+}\\
	\rhoP_{\chi_1^0}
\end{array}\right)
.
\end{equation}
Now, the generalized master equation can be divided into two parts: the first illustrating the time evolution of the occupation probabilities, Eq.~\eqref{eq:prob_def}, and the second describing the time evolution of the pseudospins, Eqs.\eqref{eq:Sigma1_def}-\eqref{eq:Sigma0_def}. 
Importantly, these equations are in general coupled to each other.

Employing Eq.~(\ref{eq:MasterEqFull}), the time evolution of the occupation probabilities can be written as
\begin{multline}
\renewcommand{\arraystretch}{1.25}
\hspace*{-10pt}
	\dfrac{\intd}{\intd t} 
	\!
	\begin{pmatrix}
	\rhoP_0 \\ \rhoP_\og \\ \rhoP_\oe
	\end{pmatrix}		
	=  
	\begin{pmatrix}
	-1/\tau_0  & \trW_{\og(0)}^- & \trW_{\oe(0)}^-
	\\  
	\trW_{\og(0)}^+ & -1/\tau_\og & 0
	\\
	\trW_{\oe(0)}^+/2 & 0 & -1/\tau_\oe
	\end{pmatrix}
	\!\!
	\cdot
	\!\!
	\begin{pmatrix}
	\rhoP_0 \\ \rhoP_\og \\ \rhoP_\oe
	\end{pmatrix}
\\
\renewcommand{\arraystretch}{1.25}
\hspace*{16pt}
	+
	2 
	\begin{pmatrix}
	\trW_{\coh(0)}^- & 0 &  0 \\
	2\trW_{\coh(\og)}^- & 0 &  0 \\
	0 & 0 & 0
	\end{pmatrix}
	\!\!
	\cdot
	\!
	\vSigmaog 
	+	
	\begin{pmatrix}
	2\trW_{0(z)}^+  \\ 2\trW_{\og(z)}^+ \\ \trW_{\oe(z)}^+
	\end{pmatrix}
	\!
	\Sigma_0^z
	\ .
	\!\!
\end{multline}	
The explicit combined expressions for elements of the kernel~`$W$'  are given in Appendix~\ref{app:exp}. Furthermore, we identify the characteristic relaxation time scales of the neutral, and charged ground/excited state as 
\begin{subequations}
\begin{eqnarray}
\label{eq:tau0}
	\tau_0
	& = &
	\frac{2\hbar}{\Gamma}
	\Big\{
	\sum_{r}
	\Big[
	f_r^+\big(\Delta_{\chi_1\chi_0}\big)
	+
	\tfrac{1}{2}
	f_r^+\big(\Delta_{\chi_1^0\chi_0}\big)
	\Big]
	\Big\}^{-1}
	\hspace*{20pt}
\\
\label{eq:tau1g}
	\tau_\og
	& = &
	\frac{2\hbar}{\Gamma}
	\Big\{
	\sum_{r}
	f_r^-\big(\Delta_{\chi_1\chi_0}\big)
	\Big\}^{-1}\\
\label{eq:tau1e}
	\tau_\oe
	& = & 
	\frac{2\hbar}{\Gamma}
	\Big\{
	\sum_{r}
	f_r^-\big(\Delta_{\chi_1^0\chi_0^{\phantom{0}}}\big)
	\Big\}^{-1}\ .
\end{eqnarray}
\end{subequations}
In the equations above, we have introduced the auxiliary notation for the Fermi functions \mbox{$f_r^-(\en)\equiv 1-f_r\big(\en\big)$} and \mbox{$f_r^+(\en)\equiv f_r\big(\en\big)$}, and   energy differences \mbox{$\Delta_{\chi\chi^\prime}\equiv\en_\chi-\en_{\chi^\prime}$}, together with the energies associated to the different spin states \mbox{$\en_{\chi_0}\equiv\en_{\chi_0^+}=\en_{\chi_0^-}$}  and~\mbox{$\en_{\chi_1}\equiv\en_{\chi_1^+}=\en_{\chi_1^-}$}. 

We find the Bloch equation for the pseudospin~$\vSigmaog$,  
\begin{align}\label{eq:vSigma1_def}
\hspace*{-4pt}
	\frac{\intd}{\intd t}\vSigmaog
	=\ &
	\renewcommand{\arraystretch}{1.25}
	-
	\frac{1}{\tau_\og}
	\vSigmaog
	+
	\vSigmaog\times\magB
	+
	\begin{pmatrix}
	\trW_{\coh(z)}^+
	\\
	0
	\\
	0
	\end{pmatrix}
	\!
	\Sigma_0^z
\nonumber\\
	&
	\renewcommand{\arraystretch}{1.25}
	+
	\frac{1}{2}
	\begin{pmatrix}
	\trW_{\coh(0)}^+
	\\
	0
	\\
	0
	\end{pmatrix}
	\!
	\rhoP_0
	+
	\begin{pmatrix}
	\trW_{\coh(\og)}^-
	\\
	0
	\\
	0
	\end{pmatrix}
	\!
	\rhoPog
	,
	\!\!
\end{align}
where the first term represents the relaxation of the~$x$,~$y$ and~$ z $-components of the pseudospin~$\vSigmaog$, with the time constant~$\tau_\og$.
The three terms involving~$\rhoP_0$, $\rhoP_\og$ and~$\Sigma_0^z$ act as source terms for the pseudospin~$\vSigmaog$. Furthermore, the term~$\vSigmaog\times\magB$ in Eq.~(\ref{eq:vSigma1_def}) describes the rotation of the pseudospin~$\vSigmaog$ around an effective magnetic 
field~\mbox{$
	\magB
	=
	(\mathcal{B}_x, \mathcal{B}_y, \mathcal{B}_z)^\text{T}
$,} 
whose  components have the following form: \mbox{$\mathcal{B}_y=0$,} \mbox{$\mathcal{B}_z=\big[\en_{\chi_1^+} - \en_{\chi_1^-}\big]/\hbar$,} and
\begin{equation}
	\mathcal{B}_x
	=
	\frac{1}{2\pi\hbar}
 	\sum_{r}
 	\big[\Gamma_\uparrow^r-\Gamma_\downarrow^r\big]
 	\Big[
 	\widetilde{\Psi}_r\big(\Delta_{\chi_0\chi_1}\big)	  
 	-
 	\ln\Big(\frac{E_\text{c}}{2\pi \kB T}\Big)
 	\Big]
	.
\end{equation}
Here, 
\mbox{$
	\widetilde{\Psi}_r(\en)
	\equiv
	\text{Re}\big\{\Psi\big[1/2+i(\en+\mu_r)/(2\pi \kB T)\big]\big\}
$,}
with $\Psi(\en)$ representing the digamma function, and $E_\text{c}$ being the largest (cut-off) energy scale.

In an analogous way, one can find the expression for the time evolution of~$\Sigma^z_{0}$,
\begin{align}\label{eq:Sigma0z_def}
\hspace*{-4pt}
	\dfrac{\intd}{\intd t}\Sigma_0^z
	=\ & 
	-	
	\dfrac{1}{\tau_0} \Sigma_0^z
	+
	\trW_{\coh(z)}^- \Sigmaog^x
\nonumber\\
	&
	+ 
	\frac{1}{2}
	\Big[
	\trW_{0(z)}^+ \rhoP_0
	+ 
	\trW_{\og(z)}^- \rhoP_\og
	+ 
	\trW_{\oe(z)}^-\rhoP_\oe
	\Big]
	.
	\!\!
\end{align}
These equations show that, in general, the dynamics of probabilities and coherences are coupled.
Specifically, if the neutral-state pseudospin~$\Sigma_0^z$, Eq.~\eqref{eq:Sigma0z_def} is not suppressed in the stationary limit, it gives rise to the~$x$ and~{$y$ components} of the charged-state pseudospin~$\vSigmaog$, as visible from Eq.~\eqref{eq:vSigma1_def}. 
Inspecting the explicit expressions for the combined elements of the kernel~`$W$' given in Appendix~\ref{app:exp}, we conclude, though, that this is the case  only if \mbox{$\Gamma_\uparrow\neq\Gamma_\downarrow$}, as it is realized for \emph{ferromagnetic} electrodes.

Consequently, in the limit of small transverse anisotropy, leading to \mbox{$\Delta\leqslant\Gamma$}, and spin-polarized electrodes, off-diagonal elements of the reduced density matrix are expected to contribute to the molecular dynamics.
However, for \emph{nonmagnetic} electrodes, the equations for the pseudospins simplify substantially,
\begin{gather}\label{eq:vSigma1_NM}
	\frac{\intd}{\intd t}\vSigmaog
	=
	\renewcommand{\arraystretch}{1.25}
	-
	\frac{1}{\tau_\og}
	\vSigmaog
	+
	\vSigmaog\times\begin{pmatrix}0\\0\\ \mathcal{B}_z\end{pmatrix}
	+
	\begin{pmatrix}
	\trW_{\coh(z)}^+
	\\
	0
	\\
	0
	\end{pmatrix}
	\!
	\Sigma_0^z
	\,,
\\
\label{eq:Sigma0z_NM}
	\dfrac{\intd}{\intd t}\Sigma_0^z
	=
	-	
	\dfrac{1}{\tau_0} \Sigma_0^z
	+
	\trW_{\coh(z)}^- \Sigmaog^x
	\,,
\end{gather}
and for the occupation probabilities one obtains
\begin{equation}\label{eq:prob_NM}
\hspace*{-5pt}
\renewcommand{\arraystretch}{1.25}
	\dfrac{\intd}{\intd t} 
	\!
	\begin{pmatrix}
	\rhoP_0 \\ \rhoP_\og \\ \rhoP_\oe
	\end{pmatrix}	
	\!\!=\!\! 
	\begin{pmatrix}
	-1/\tau_0  & \trW_{\og(0)}^- & \trW_{\oe(0)}^-
	\\  
	\trW_{\og(0)}^+ & -1/\tau_\og & 0
	\\
	\trW_{\oe(0)}^+/2 & 0 & -1/\tau_\oe
	\end{pmatrix}	
	\!
	\cdot
	\!
	\begin{pmatrix}
	\rhoP_0 \\ \rhoP_\og \\ \rhoP_\oe
	\end{pmatrix}
	\!\!.
\end{equation}
Importantly, one can see that the time evolution of probabilities, Eq.~(\ref{eq:prob_NM}), decouples from that for pseudospins, Eqs.~(\ref{eq:vSigma1_NM})-(\ref{eq:Sigma0z_NM}). Moreover, in the stationary limit one finds from Eqs.~(\ref{eq:vSigma1_NM})-(\ref{eq:Sigma0z_NM})  \mbox{$\Sigmaog^x=\Sigmaog^y=\Sigmaog^z=\Sigma_0^z=0$}, which basically means that the off-diagonal elements of the reduced density matrix of the molecule, Eq.~(\ref{eq:rhored_ex}), vanish.
%

\section{Composite transition rates~`$\trW$'}\label{app:exp}

The explicit expressions for elements of the kernel~`$W$' used in Appendix~\ref{app:coherences} are given by: 
\begin{equation}\label{eq:W0}
 	\trW_{0(z)}^+
	=
 	\frac{1}{2\hbar}
 	\sum_{r}
 	\big[\Gamma_\downarrow^r-\Gamma_\uparrow^r\big]
 	\Big\{
 	f_r^+\big(\Delta_{\chi_1\chi_0}\big)
 	-
 	\tfrac{1}{2}
 	f_r^+\big(\Delta_{\chi_1^0\chi_0}\big)
 	\Big\}
 	,
\end{equation}
\begin{equation}
 	\trW_{\coh(\og)}^-
	=
 	\frac{1}{2\hbar}
 	\sum_{r}
 	\big[\Gamma_\downarrow^r-\Gamma_\uparrow^r]
 	f_r^-\big(\Delta_{\chi_1\chi_0}\big)
 	,
\end{equation} 
\begin{equation}
\renewcommand{\arraystretch}{1.25}
	\begin{pmatrix}
	\trW_{\coh(0)}^\pm \\ \trW_{\coh(z)}^\pm
	\end{pmatrix}
	\!
 	=
 	\frac{1}{2\hbar}
 	\sum_{r}
 	\!
 	\begin{pmatrix}
 	\Gamma_\uparrow^r-\Gamma_\downarrow^r \\ \Gamma
 	\end{pmatrix}
 	\!
 	f_r^\pm\big(\Delta_{\chi_1\chi_0}\big)
 	,
\end{equation}
\begin{equation}
\renewcommand{\arraystretch}{1.25}
	\begin{pmatrix}
	\trW_{\og(0)}^\pm \\ \trW_{\og(z)}^\pm
	\end{pmatrix}
	\!
	=
	\frac{1}{2\hbar}
	\sum_{r\sigma}
	\!
	\begin{pmatrix}
	\Gamma \\ \Gamma_\uparrow^r-\Gamma_\downarrow^r
	\end{pmatrix}
	\!
 	f_r^\pm\big(\Delta_{\chi_1\chi_0}\big)
 	,
\end{equation}
\begin{equation}\label{eq:W1e}
\renewcommand{\arraystretch}{1.25}
	\begin{pmatrix}
	\trW_{\oe(0)}^\pm \\ \trW_{\oe(z)}^\pm
	\end{pmatrix}
	\!
	=  
	\frac{1}{2\hbar}
	\sum_{r\sigma}
 	\!
 	\begin{pmatrix}
 	\Gamma \\ \Gamma_\downarrow^r-\Gamma_\uparrow^r
 	\end{pmatrix}
 	\!
 	f_r^\pm\big(\Delta_{\chi_1^0\chi_0^{\phantom{0}}}\big)
	.
\end{equation}
Note that since we consider the limit  \mbox{$\big|\en_{\chi_1^+}-\en_{\chi_1^-}\big|\lesssim\Gamma$}, we have assumed \mbox{$\en_{\chi_1}\equiv\en_{\chi_1^+}=\en_{\chi_1^-}$} when deriving these expressions, in order to consistently include terms in leading order~$\Gamma$.



%

\end{document}